\documentclass[12pt]{article}
\usepackage{psfrag}
\usepackage{graphics}
\usepackage{amssymb,epsfig,amsmath,euscript,array}

\newcounter{multieqs}

\newcommand{\be}{\begin{equation}}
\newcommand{\ee}{\end{equation}}
\newcommand{\eq}[1]{(\ref{#1})}

\newcommand{\ket}[1]{|#1 \rangle}

\newcommand{\bm}[1]{\mbox{\boldmath $#1$}}

\def\bd{\begin{document}}
\def\ed{\end{document}}
\def\nn{\nonumber}
\def\bea{\begin{eqnarray}}
\def\eea{\end{eqnarray}}
\let\bm=\bibitem
\let\la=\label

\def\npb#1#2#3{Nucl. Phys. {\bf{B#1}} #3 (#2)}
\def\plb#1#2#3{Phys. Lett. {\bf{#1B}} #3 (#2)}
\def\prl#1#2#3{Phys. Rev. Lett. {\bf{#1}} #3 (#2)}
\def\prd#1#2#3{Phys. Rev. {D \bf{#1}} #3 (#2)}
\def\cmp#1#2#3{Comm. Math. Phys. {\bf{#1}} #3 (#2)}
\def\cqg#1#2#3{Class. Quantum Grav. {\bf{#1}} #3 (#2)}
\def\nppsa#1#2#3{Nucl. Phys. B (Proc. Suppl.) {\bf{#1A}}#3 (#2)}
\def\ap#1#2#3{Ann. of Phys. {\bf{#1}} #3 (#2)}
\def\ijmp#1#2#3{Int. J. Mod. Phys. {\bf{A#1}} #3 (#2)}
\def\rmp#1#2#3{Rev. Mod. Phys. {\bf{#1}} #3 (#2)}
\def\mpla#1#2#3{Mod. Phys. Lett. {\bf A#1} #3 (#2)}
\def\jhep#1#2#3{J. High Energy Phys. {\bf #1} #3 (#2)}
\def\atmp#1#2#3{Adv. Theor. Math. Phys. {\bf #1} #3 (#2)}

\newcommand{\EQ}[1]{\begin{equation} #1 \end{equation}}
\newcommand{\AL}[1]{\begin{subequations}\begin{align} #1
\end{align}\end{subequations}}
\newcommand{\SP}[1]{\begin{equation}\begin{split} #1 \end{split}\end{equation}}
\newcommand{\ALAT}[2]{\begin{subequations}\begin{alignat}{#1} #2
\end{alignat}\end{subequations}}
\def\beqa{\begin{eqnarray}}
\def\eeqa{\end{eqnarray}}
\def\beq{\begin{equation}}
\def\eeq{\end{equation}}

\def\N{{\cal N}}
\def\sst{\scriptscriptstyle}
\def\thetabar{\bar\theta}
\def\Tr{{\rm Tr}}
\def\one{\mbox{1 \kern-.59em {\rm l}}}

\def\a{\alpha}      \def\da{{\dot\alpha}}
\def\b{\beta}       \def\db{{\dot\beta}}
\def\c{\gamma}  \def\C{\Gamma}  \def\cdt{\dot\gamma}
\def\d{\delta}  \def\D{\Delta}  \def\ddt{\dot\delta}
\def\e{\epsilon}        \def\vare{\varepsilon}
\def\f{\phi}    \def\F{\Phi}    \def\vvf{\f}
\def\h{\eta}
\def\k{\kappa}
\def\l{\lambda} \def\L{\Lambda}
\def\m{\mu} \def\n{\nu}
\def\o{\omega}
\def\p{\pi} \def\P{\Pi}
\def\r{\rho}
\def\s{\sigma}  \def\S{\Sigma}
\def\t{\tau}
\def\th{\theta} \def\Th{\Theta} \def\vth{\vartheta}
\def\X{\Xeta}
\def\z{\zeta}

\def\cA{{\cal A}} \def\cB{{\cal B}} \def\cC{{\cal C}}
\def\cD{{\cal D}} \def\cE{{\cal E}} \def\cF{{\cal F}}
\def\cG{{\cal G}} \def\cH{{\cal H}} \def\cI{{\cal I}}
\def\cJ{{\cal J}} \def\cK{{\cal K}} \def\cL{{\cal L}}
\def\cM{{\cal M}} \def\cN{{\cal N}} \def\cO{{\cal O}}
\def\cP{{\cal P}} \def\cQ{{\cal Q}} \def\cR{{\cal R}}
\def\cS{{\cal S}} \def\cT{{\cal T}} \def\cU{{\cal U}}
\def\cV{{\cal V}} \def\cW{{\cal W}} \def\cX{{\cal X}}
\def\cY{{\cal Y}} \def\cZ{{\cal Z}}

\def\ua{\underline{\alpha}}
\def\ub{\underline{\phantom{\alpha}}\!\!\!\beta}
\def\uc{\underline{\phantom{\alpha}}\!\!\!\gamma}
\def\um{\underline{\mu}}
\def\ud{\underline\delta}
\def\ue{\underline\epsilon}
\def\una{\underline a}\def\unA{\underline A}
\def\unb{\underline b}\def\unB{\underline B}
\def\unc{\underline c}\def\unC{\underline C}
\def\und{\underline d}\def\unD{\underline D}
\def\une{\underline e}\def\unE{\underline E}
\def\unf{\underline{\phantom{e}}\!\!\!\! f}\def\unF{\underline F}
\def\unm{\underline m}\def\unM{\underline M}
\def\unn{\underline n}\def\unN{\underline N}
\def\unp{\underline{\phantom{a}}\!\!\! p}\def\unP{\underline P}
\def\unq{\underline{\phantom{a}}\!\!\! q}
\def\unQ{\underline{\phantom{A}}\!\!\!\! Q}
\def\unH{\underline{H}}

\def\As {{A \hspace{-6.4pt} \slash}\;}
\def\bs {{b \hspace{-6.4pt} \slash}\;}
\def\Ds {{D \hspace{-6.4pt} \slash}\;}
\def\ds {{\del \hspace{-6.4pt} \slash}\;}
\def\ss {{\s \hspace{-6.4pt} \slash}\;}
\def\ks {{ k \hspace{-6.4pt} \slash}\;}
\def\ps {{p \hspace{-6.4pt} \slash}\;}
\def\pas {{{p_1} \hspace{-6.4pt} \slash}\;}
\def\pbs {{{p_2} \hspace{-6.4pt} \slash}\;}

\def\Fh{\hat{F}}
\def\Nh{\hat{N}}
\def\Vh{\hat{V}}
\def\Xh{\hat{X}}
\def\ah{\hat{a}}
\def\xh{\hat{x}}
\def\yh{\hat{y}}
\def\ph{\hat{p}}
\def\xih{\hat{\xi}}

\def\psit{\tilde{\psi}}
\def\Psit{\tilde{\Psi}}
\def\tht{\tilde{\th}}

\def\At{\tilde{A}}
\def\Qt{\tilde{Q}}
\def\Rt{\tilde{R}}
\def\Nt{\tilde{N}}

\def\at{\tilde{a}}
\def\st{\tilde{s}}
\def\ft{\tilde{f}}
\def\pt{\tilde{p}}
\def\qt{\tilde{q}}
\def\vt{\tilde{v}}
\def\nt{\tilde{n}}

\def\delb{\bar{\partial}}
\def\bz{\bar{z}}
\def\bD{\bar{D}}
\def\bB{\bar{B}}

\def\bk{{\bf k}}
\def\bl{{\bf l}}
\def\bp{{\bf p}}
\def\bq{{\bf q}}
\def\br{{\bf r}}
\def\bx{{\bf x}}
\def\by{{\bf y}}
\def\bR{{\bf R}}
\def\bV{{\bf V}}

\def\d{\delta}\def\D{\Delta}\def\ddt{\dot\delta}

\def\pa{\partial} \def\del{\partial}
\def\xx{\times}
\def\uno{\mbox{1 \kern-.59em {\rm l}}}

\def\trp{^{\top}}
\def\inv{^{-1}}
\def\dag{{^{\dagger}}}
\def\pr{^{\prime}}

\def\rar{\rightarrow}
\def\lar{\leftarrow}
\def\lrar{\leftrightarrow}

\newcommand{\0}{\,\!}      
\def\one{1\!\!1\,\,}
\def\im{\imath}
\def\jm{\jmath}

\newcommand{\tr}{\mbox{tr}}
\newcommand{\slsh}[1]{/ \!\!\!\! #1}

\def\vac{|0\rangle}
\def\lvac{\langle 0|}

\def\hlf{\frac{1}{2}}
\def\ove#1{\frac{1}{#1}}

\def\Box{\square}
\def\ZZ{\mathbb{Z}}
\def\CC#1{({\bf #1})}
\def\bcomment#1{}
\def\bfhat#1{{\bf \hat{#1}}}
\def\VEV#1{\left\langle #1\right\rangle}

\newcommand{\ex}[1]{{\rm e}^{#1}} \def\ii{{\rm i}}

\def\rr{{\rm r}} \def\rs{{\rm s}}\def\rv{{\rm v}}
\def\ri{{\rm i}}\def\rj{{\rm j}}

\newcommand{\lrbrk}[1]{\left(#1\right)}
\newcommand{\sfrac}[2]{{\textstyle\frac{#1}{#2}}}

\begin{document}

\hfill{hep-th/0302064}

\vspace{20pt}

\begin{center}

{\Large \bf BMN operators with three scalar impurites\\}
\vspace{10pt}
{\Large \bf  and the vertex--correlator duality in pp-wave}

\vspace{30pt}

{\bf George Georgiou and Valentin V. Khoze}

{\small \em
Centre for Particle Theory,
Department of Physics and IPPP,\\
University of Durham, Durham, DH1 3LE, UK
}

\vspace{10pt}

Email: {\sffamily \tt 
george.georgiou@durham.ac.uk, valya.khoze@durham.ac.uk }

\vspace{30pt} {\bf Abstract}

\end{center}

We calculate 3-point correlation functions of $\Delta$-BMN operators
with 3 scalar impurites in $\cN=4$ supersymmetric gauge theory.
We use these results to test the pp-wave/SYM duality correspondence of the 
{\it vertex--correlator} type. This correspondence relates the coefficients of
3-point correlators of $\Delta$-BMN operators in gauge theory to the 3-string 
vertex in lightcone string field theory in the pp-wave background.
We verify the vertex--correlator duality equation of hep-th/0301036 at 
the 3 scalar impurites level for supergravity and for string modes.

\vspace{0.5cm}

\setcounter{page}0
\thispagestyle{empty}
\newpage


\section{Introduction}

This paper continues the study of correlation functions of BMN operators
\cite{Constable1,CKT,n-point,BKPSS,Constable2,CK} in the light of the
pp-wave/SYM correspondence \cite{BMN}.

The pp-wave/SYM correspondence of Berenstein, Maldacena and Nastase (BMN) 
\cite{BMN}
represents all massive modes of type IIB superstring in the plane wave 
background
in terms of composite BMN operators of $\cN=4$ Yang-Mills theory in 4D.
In its minimal form, this correspondence emphasizes a duality relation 
between the masses of string states and the anomalous dimensions of the
corresponding BMN operators in gauge theory in the large $N$ double 
scaling limit. 
This relation has been verified
in the planar limit of SYM perturbation 
theory 
in \cite{BMN,gross,zanon}. 
Calculations in the BMN sector of gauge theory at the
nonplanar level were performed in
\cite{KPSS,Constable1,BKPSS,Constable2} also taking into account
mixing effects of planar BMN operators.
The minimal mass--dimension type duality relation was extended
in \cite{ver,gross2,bits2}
to all orders in the effective genus expansion parameter $g_2$
and expressed in the form
${H}_{\rm string} = {H}_{\rm SYM} - J.$
Here ${H}_{\rm string}$ is the full string field 
theory Hamiltonian, and ${H}_{\rm SYM} - J = \Delta-J$ is 
the gauge theory Hamiltonian
(the conformal dimension) minus the R-charge. Recent work in this 
direction
includes \cite{gomis}.

In this paper, instead, we address a more ambitious duality relation 
\cite{CK} of a 
{\it vertex--correlator} type, summarized in the next Section.
This type of correspondence for pp-waves
was first discussed in \cite{Constable1} and 
relates the coefficients of 3-point correlators of BMN operators
in gauge theory to 3-string vertices in lightcone string field theory in the
pp-wave background.
It is well-known that 
in the AdS/CFT scenario, in addition to the relation
between the masses of supergravity states and the 
dimensions of the dual gauge theory operators, 
one can also compare directly the
correlation functions in gauge theory with supergravity interactions
in the bulk \cite{witten,gkp}. 
Since the pp-wave/CFT correspondence can
be viewed as a particular limit of the AdS/CFT correspondence, 
it is natural to expect that a version of vertex--correlator
type duality will hold in the pp-wave/SYM correspondence.

Building on previous work \cite{Constable1,huang,CKT,BKPSS},
the authors of \cite{CK} were able to represent all known gauge theory
results for 3-point functions of BMN operators with 2 scalar impurites
in terms of a single concise expression involving the 3-string vertex
in light-cone string field theory in the pp-wave background. 
The goal of the
present paper is to test this relation at the level of BMN operators with
3 scalar impurites. 

In conformal theory, the two- and three-point functions of conformal primary 
operators are completely determined by conformal invariance of the
theory. One can always
choose a basis of scalar conformal primary operators such that the 
two-point functions
take the canonical form:
\begin{equation} \label{2pt}
\langle {\cO}_I (x) \bar\cO_J(0) \rangle = \frac{\d_{IJ}}{(4 \pi^2
x^2)^{\Delta_I}},
\end{equation}
and all the nontrivial information of the
three-point function is contained in the $x$-independent 
coefficient $C_{1 2 3}$:
\begin{equation} \label{3pt}
\langle \cO_{1}(x_1) \cO_{2}(x_2) \bar\cO_{3}(0) \rangle  =
\frac{C_{123} }
{(4\pi^2 x_{12}^2)^{\frac{\D_1+\D_2 -\D_3}{2}}
 (4\pi^2 x_{1}^2)^{\frac{\D_1+\D_3 -\D_2}{2}}
 (4\pi^2 x_{2}^2)^{\frac{\D_2+\D_3 -\D_1}{2}}},
\end{equation}
where $x_{12}^2: = (x_1-x_2)^2$.
Since the form of the $x$-dependence of conformal 3-point functions is universal,
it is natural to expect that $C_{123}$
is related to the interaction of the corresponding three string states
in  the pp-wave background.
Note, that in order to be able to use the coefficients $C_{123}$,
it is essential to work on the SYM side with $\Delta$-BMN operators.
These operators are defined in such a way that they do not mix 
with each other (i.e. have definite scaling
dimensions $\Delta$) and which are conformal primary operators. 
Conformal invariance
of the $\cN=4$ theory then implies that the 2-point correlators 
of scalar $\Delta$-BMN 
operators are canonically normalized, 
and the 3-point functions take the simple form \eqref{3pt}. 
Defined in this way, the basis of $\Delta$-BMN operators is unique and distinct 
from other BMN bases considered in the literature. For 2 scalar impuritites,
this $\Delta$-BMN basis was constructed in \cite{BKPSS}.

In this paper we will work with scalar\footnote{Vector 
operators, i.e. $\Delta$-BMN operators with vector
impurites, discussed in \cite{gursoy,beisert,klose}, will be considered elsewhere.}
$\Delta$-BMN operators $\cO_{n_{1} n_{2}n_{3}}$
with 3 impurites which correspond to 
the string bra-state $\langle 0 | \alpha^1_{n_1}\alpha^2_{n_2}\alpha^3_{n_3}$.
In string theory 
$n_{i}$ are the labels of string oscillators $\alpha^i_{n_i}$, 
and the level matching
constraint is $n_1+n_2+n_3=0$. 
Bare single-trace BMN operators  with 3 different real scalar impurities 
\cite{Constable1,gross2} are given by
\begin{eqnarray} 
\cO_{n_{1}n_{2}n_{3}} \equiv \cO^{123}_{n_{1}n_{2}n_{3}} + 
\cO^{132}_{n_{1}n_{2}n_{3}} 
=\, \frac{1}{J\sqrt{N^{J+3}}}\sum_{0\leq l,k}^{l+k\leq {J}}[ q_{2}^{l}q_{3}^{l+k} \tr(\phi_{1}Z^{l}\phi_{2}Z^{k}\phi_{3}Z^{J-l-k})\nonumber\\+ q_{3}^{l}q_{2}^{l+k} \tr(\phi_{1}Z^{l}\phi_{3}Z^{k}\phi_{2}Z^{J-l-k})]
\label{3imop}
\end{eqnarray}
where $q_{2}=e^{2\pi i n_{2}/J}$ and $q_{3}=e^{2\pi i n_{3}/J}$, and from now on,
we always set $n_1=-n_2-n_3$. There are two terms on the right hand side
of \eqref{3imop} since there are two inequivalent orderings of $\phi_{1}$,
$\phi_{2}$ and $\phi_{3}$ inside the trace\footnote{Note that in both terms
$\phi_2$ in the $Z$-position $l$ is accompanied by $q_2^l$, similarly for $\phi_3$.
Hence, each of the two terms contributes to the same string state.}.
The pp-wave/SYM duality is supposed to hold in the BMN large $N$ 
double scaling limit,
\be \label{doublel}
J \sim \sqrt{N} \ , \qquad N\to \infty .
\ee 
In this limit there remain two free finite dimensionless
parameters \cite{BMN,KPSS,Constable1}: 
the effective coupling constant of the BMN sector of gauge theory,
\be \label{lampr}
 \l' = \frac{g_{\rm YM}^2 N}{J^2} = \frac{1}{(\mu p^+ \a')^2}\ 
\ee
and the effective genus counting parameter
\begin{equation} \label{gtwo}
g_2 :=\frac{J^2}{N}= 4 \pi g_s (\mu p^+ \a')^2 \ .
\end{equation} 
The right hand sides of \eqref{lampr},\eqref{gtwo} express $\l'$ and $g_2$ in terms
of pp-wave string theory parameters.

Operators \eqref{3imop} are the starting point for building the $\Delta$-BMN operators.
In interacting field theory, bare operators have to be UV-renormalized and
the effects of operator mixing have to be taken into account. It is well-known
by now \cite{arut,bianchi,BKPSS,Constable2}
that the single-trace BMN operators mix with the multi-trace operators
even in free theory at non-planar level, i.e. starting from order $g_2 (\l')^0$.
Hence, in order to calculate the leading-order contribution to the 3-point coefficient
$C_{123}\propto g_2$ in \eqref{3pt}, one has to work with the order $g_2$
$\Delta$-BMN operators which involve the single-trace expressions \eqref{3imop}
plus a linear combination of double-trace operators with coefficients of 
order $g_2 (\l')^0$.
For the simpler case of 2 scalar impurites,
the single-double trace mixing effects have been calculated in \cite{BKPSS}, and the
corresponding conformal 3-point function coefficients $C_{123}$ were determined. 
One of the main technical results of the present paper will be a determination
of the coefficients $C_{123}$ for $\Delta$-BMN operators with 3 scalar impurities
(and general oscillator labels $n_2,n_3 \in Z$).

The paper is organized as follows.
In Section 2 we summarize the vertex--correlator duality proposal
of \cite{CK} and write down the relevant equations.
In Section 3 we calculate 2-point correlators of operators $\cO_{n_{1} n_{2}n_{3}}$
to order $\l'$ in planar perturbation theory in the BMN limit. This
is necessary in order to canonically normalize the UV-renormalized operators
to order $\l'$. 
Section 4 contains our main technical results on the field theory side. 
There we calculate 3-point functions involving operators  $\cO_{n_{1} n_{2}n_{3}}$.
We first derive the conformal expression \eqref{3pt} and extract the coefficient
$C_{123}$ for 3-point functions containing 1 general and 2 chiral operators
(i.e. 1 string and 2 supergravity states in dual string theory).
We then generalize this calculation of $C_{123}$ to the case of two string states.
In the final Section we demonstrate that the results of Section 4 are in complete
agreement with the vertex--correlator duality prediction \cite{CK} of Section 2.

\section{The vertex--correlator duality}

Here we give a brief summary of the duality relation.
For more detail, we refer the reader to
\cite{CK}.

For the bosonic external string states $\langle \Phi_i|$
the proposed correspondence relation is
\be \label{hhhh}
\mu(\Delta_1+\Delta_2-\Delta_3)\, C_{123}
= 
\langle \Phi_1| \langle \Phi_2|\langle \Phi_3| \, {\sf P}\, 
\exp( \frac{1}{2} \sum_{\rr,\rs=1}^3 \sum_{m,n=-\infty}^\infty
\sum_{I=1}^8 \a^{\rr\, I\dag}_{m} \Nh_{mn}^{\rr\rs}  
\a^{\rs\, I\dag}_{n}) \,\ket{0}_{123}.
\ee 
This relation, is conjectured to be valid to all orders in $\l'$ and to the
leading order in $g_2$ in the double scaling limit \eqref{doublel}.
Equation \eqref{hhhh},
originally proposed in \cite{Constable1}, is the first
key element of the vertex-correlator duality.
$\Nh_{mn}$ are the Neumann matrices in the $\alpha$-basis
of string oscillators.
These matrices were recently calculated in \cite{HSSV} as an 
expansion in inverse powers of $\mu$ at $\mu\to \infty$.
Results of \cite{HSSV} for $\Nh_{mn}$
constitute the second element of the proposed duality.
The relevant for us leading order expressions of $\Nh_{mn}$ directly in the
$\alpha$-oscillator basis can be found
in the the Appendix.

The third and final element of the vertex--correlator duality is the 
expression \cite{CK} for the bosonic part of
the string field theory prefactor, ${\sf P}$, 
which appears on the right hand side of \eqref{hhhh},
\be \label{pf}
{\sf P} =\, (-1)^p\,
C_{\rm norm}\,({\sf P}_{I}+{\sf P}_{II}) \ ,
\ee
where\footnote{We are using standard definitions for the SFT 
quantities in the pp-wave background such as 
$\omega_{\rr m}$, $\alpha_\rr$ and $\mu$, which are summarized in the Appendix.}
\be \label{pf1}
{\sf P}_{I} = 
\sum_{\rr=1}^3\sum_{m=-\infty}^{+\infty}\frac{\omega_{\rr m}}{\alpha_\rr}\, 
\a^{\rr \, I\dag}_m \a^{\rr\, I}_m \ ,
\ee
\be \label{pf2}
{\sf P}_{II} = \frac{1}{2}
\sum_{\rr,\rs=1}^{3}\sum_{m,n > 0}
\frac{\omega_{\rr m}}{\alpha_\rr}\,
(\hat{N}^{\rr\rs}_{m -n} - \hat{N}^{\rr\rs}_{m n})
(\a^{\rr\, I \dag}_m \a^{\rs\, I\dag}_n + 
\a^{\rr\, I \dag}_{-m} \a^{\rs\, I \dag}_{-n} - 
\a^{\rr\, I \dag}_{m} \a^{\rs\, I \dag}_{-n} - \a^{\rr\, I \dag}_{-m} 
\a^{\rs\, I\dag}_{n} )
\ee
and
\be \label{cnorm}
C_{\rm norm} = g_2 \frac{\sqrt{y(1-y)}}{\sqrt{J}}= C^{\rm vac}_{123} \ .
\ee
The only new ingredient here compared to \cite{CK} is the overall sign $(-1)^p$
in \eqref{pf}, where $p\equiv \frac{1}{2} \sum_{\rr=1}^3\sum_{m=-\infty}^{+\infty} 
\a^{\rr \, I\dag}_m \a^{\rr\, I}_m$ counts the number of impurities.
For all the cases involving BMN operators with 2 impurities considered in \cite{CK}, 
it turns out that $(-1)^p=(-1)^2=1$, and hence is irrelevant.
In the present paper, all the cases involving 3 impurities will lead to an overall
minus sign,  $(-1)^p=(-1)^3=-1$.

In terms of the original SFT $a$-oscillator basis the full prefactor takes a 
remarkably simple form
\be
{\sf P} =\, (-1)^p\,C_{\rm norm}  \sum_{\rr =1}^{3} \left(
\sum_{m > 0} 
\frac{\omega_{\rr m}}{\alpha_\rr}\,
a^{\rr\, I \dag}_{m} a^{\rr\, I}_{m}
+ \mu \, \mbox{sign} (\a_\rr) a_0^{\rr\, I \dag} a_0^{\rr\, I}
\right) ,
\ee
however, as in \cite{CK}, we will continue using the prefactor 
in the BMN $\a$-oscillator basis, \eq{pf1} and \eq{pf2}, where the
comparison with the gauge theory BMN correlators is most direct.

This prefactor, and in particular the second term ${\sf P}_{II}$, 
was constructed in \cite{CK} to 
reproduce a particular class of field theory results
for the 3-point functions\footnote{
A first principles
derivation of the string field theory prefactor is highly
desirable, but (at least in our view) not yet available
inspite of much progress made
in the pp-wave lightcone string field theory
\cite{SV1,SV2,rel3,ari1,ari2,z2-1,z2-2,HSSV}.
We note that \eqref{pf} is different from the earlier proposals for the prefactor 
in \cite{SV2,ari1,HSSV}.}.
 It was then successfully tested 
in \cite{CK} against all the available field theory results involving
BMN operators with 2 scalar impurites and also the simplest cases
involving BMN operators with 3 impurities.
In Section 5 we will verify that the duality relation \eqref{hhhh}
with the prefactor  \eqref{pf} holds at the 3-impurity level.

\medskip

\centerline{**********}

We emphasize that the
matching to field theory results is highly non-trivial even though
the choice \cite{CK} of the
prefactor in \eqref{pf} is ``phenomenological''. In the next
two Sections we will assemble a detailed SYM calculation of the
3-point coefficients for BMN correlators with 3 impurities, \eqref{gdef},
\eqref{gpdef}. This calculation is presented in detail
in order to convince
the reader that a {\it coincidental} agreement of our SYM results and the 
string vertex with the prefactor \eq{pf} 
(which a priory knows nothing about 3-impurity operators) is
very unlikely. 
The reader primarily interested in the tests of the correspondence,
can skip directly to the final SYM results,
Eqs. \eqref{cst123}, \eqref{c0123} and \eqref{ctil123}, \eqref{ctil0}
and then to the last Section.

\section{Two-point correlators}

As explained earlier, on the SYM side of our proposed correspondence 
we must use the $\Delta$-BMN operators $\cO$. For BMN operators with 
2 scalar impurities this basis was
constructed in \cite{BKPSS} to order ${g_2} (\l')^0$ and
${g_2}^2(\l')^0$ and involves a linear combination
of the original single-trace BMN operator and the double-trace 
(in general multi-trace)
BMN operators.

There are two important cases where simplifications occur such that at the leading 
non-vanishing order in $g_2$, only the single-trace operators \eqref{3imop} need
to be taken into account. The first case involves 2-point functions
$\langle \cO \bar{\cO} \rangle$, and will be considered in this Section.
The second case involves
3-point functions $\langle \cO_1 \cO_2 \bar{\cO}_3 \rangle$, where 
$\cO_1$ and $\cO_2$ are chiral BMN operators, and $\cO_3$ is a general one, 
i.e. two supergravity and one string state in dual string theory. 
This case will be considered in the first part of the next Section.
It is easy to check (see e.g. \cite{BKPSS})
that in both cases the contributions from 
double- and higher trace operators to $\cO$'s give vanishing contributions
to the correlators at the leading order in
$g_2$ and in the double-scaling limit \eqref{doublel}.

Before we continue we make a final general comment. 
One can split scalar interactions of the $\cN=4$ SYM 
Lagrangian into D-terms, F-terms and K-terms as is done in 
\cite{Constable1,BKPSS} and show that at one loop level 
the  D-terms cancel against the gluon exchanges 
and scalar self-energies. So one is left only with  
F-terms and K-terms. However the K-terms have vanishing 
contribution in the cases we are going to consider since  
K-terms couple only to SO(6) traces.
Thus, there is only an F-term interaction to consider
which has a factor of $g_{YM}^{2}$ for every 
vertex where a $\phi^{i}$ line crosses a 
$Z=\frac{\phi^{5}+i \phi^{6}}{\sqrt{2}}$ line,
and a factor of $-g_{YM}^{2}$ when the lines 
do not cross.

In this Section we calculate 2-point correlators $\langle \cO \bar{\cO} \rangle$
of renormalized operators \eqref{3imop} in planar perturbation theory
to order $\l'$. This is needed to normalize the operators correctly,
such that \eqref{2pt} holds at order $\l'$. In this and the next Section we 
will be calculating Feynman diagrams 
in dimensional reduction to $4-2\e$ dimensions, and
in coordinate space. Our calculations follow and generalize the approach of \cite{CKT}.

We note that bare operators in \eqref{3imop} were normalized in such a way
that their free 2-point planar correlator is 
\begin{eqnarray}
\langle\bar{\cO}_{n_{1} n_{2}{n}_{3}}(0)
\cO_{\tilde{n}_{1}\tilde{n}_{2}\tilde{n}_{3}}(x)\rangle=
\delta_{{n}_{2},\tilde{n}_{2}}\delta_{n_{3},\tilde{n}_{3}}\Delta(x)^{J+3} 
\end{eqnarray}
in the BMN limit \eqref{doublel}. 
Here $ \Delta(x)$ is the scalar propagator,
\be
\Delta(x)=\frac{\Gamma(1-\e)}{ (4 \pi^{2-\e})(x^{2})^{1-\e}} 
\ee
There are four contributions to consider, 
$\langle\bar{\cO}^{123}(0) \cO^{123}(x)\rangle$,
$\langle\bar{\cO}^{132}(0) \cO^{132}(x)\rangle$, 
$\langle\bar{\cO}^{123}(0) \cO^{132}(x)\rangle$ and
$\langle\bar{\cO}^{132}(0) \cO^{123}(x)\rangle$. 
The last two correlators vanish in free theory (since the 3 $\phi$'s are different),
and will be shown to vanish also at order $\l'$ at the planar level.

We first calculate the interacting part of 
$\langle\bar{\cO}^{123}(0) \cO^{123}(x)\rangle$, 
\be \label{ottott}
\langle\bar{\cO}^{123}(0) \cO^{123}(x)\rangle
 =\frac{\Delta(x)^{J+3}}{J^{2}}(-g_{YM}^{2}N)\, I(x)\, (P_{1}+P_{2}+P_{3})
\ee
where $I(x)$ is the interaction integral
with $\D(x)^2$ removed: 
\bea \label{Ix}
I(x) &=& \left(\frac{\C(1-\e)}{4 \pi^{2-\e}}\right)^2 (x^2)^{2-2\e}
\int \frac{d^{4-2\e}y}{(y^2)^{2-2\e} (y-x)^{2(2-2\e)}}\nn\\
&=& \frac{1}{8\pi^2} (\frac{1}{\e} + \c +1 + \log \pi + \log x^2 + O(\e)).
\eea
We will use a subtraction scheme which subtracts the $1/\e$ pole together
with (an arbitrary) finite part $s$
\begin{equation}  \label{sub}
\frac{1}{\e} + s.
\end{equation}
$P_{1}$, $P_{2}$ and $P_{3}$ on the right hand side of \eqref{ottott}
are the total contributions of the phase factors for the diagrams
of Figure 1, Figure 2 and Figure 3 respectively. 
Denoting by $q_{2}$, $q_{3}$ the BMN phases of the 
$\cO_{\tilde{n}_{1}\tilde{n}_{2}\tilde{n}_{3}} (x)$ operator, and by
$\mbox{\={r}}_{2}, \mbox{\={r}}_{3} $ the BMN phases of the 
$\bar{\cO}_{n_{1}n_{2}{n}_{3}} (0)$ operator, we obtain
\begin{eqnarray}
P_{1}=\sum_{1\leq l,0\leq k}^{l+k\leq{J}}[ q_{2}^{l}q_{3}^{l+k}
\mbox{\={r}}_{2}^{l}\mbox{\={r}}_{3}^{l+k}-
q_{2}^{l}q_{3}^{l+k}\mbox{\={r}}_{2}^{l-1}\mbox{\={r}}_{3}^{l+k-1}]\nonumber\\
+\sum_{0\leq l,0\leq k}^{l+k\leq {J-1}}[ q_{2}^{l}q_{3}^{l+k}
\mbox{\={r}}_{2}^{l}\mbox{\={r}}_{3}^{l+k}-
q_{2}^{l}q_{3}^{l+k}\mbox{\={r}}_{2}^{l+1}\mbox{\={r}}_{3}^{l+k+1}]\nonumber\\
=\sum_{0\leq l,0\leq k}^{l+k\leq {J-1}}[ q_{2}^{l+1}
q_{3}^{l+k+1}\mbox{\={r}}_{2}^{l+1}
\mbox{\={r}}_{3}^{l+k+1}-q_{2}^{l+1}q_{3}^{l+k+1}
\mbox{\={r}}_{2}^{l}\mbox{\={r}}_{3}^{l+k}]\nonumber\\
+\sum_{0\leq l,0\leq k}^{l+k\leq {J-1}}[ q_{2}^{l}q_{3}^{l+k}
\mbox{\={r}}_{2}^{l}\mbox{\={r}}_{3}^{l+k}-q_{2}^{l}q_{3}^{l+k}
\mbox{\={r}}_{2}^{l+1}\mbox{\={r}}_{3}^{l+k+1}]\nonumber\\
=\sum_{0\leq l,0\leq k}^{l+k\leq {J-1}} q_{2}^{l}q_{3}^{l+k}
\mbox{\={r}}_{2}^{l}\mbox{\={r}}_{3}^{l+k}(1+ q_{2}q_{3}
\mbox{\={r}}_{2}\mbox{\={r}}_{3}- q_{2}q_{3}-\mbox{\={r}}_{2}\mbox{\={r}}_{3})
\nonumber\\
=\sum_{0\leq l,0\leq k}^{l+k\leq {J-1}}(q_{2}\mbox{\={r}}_{2})^{l}(q_{3}
\mbox{\={r}}_{3})^{l+k}[(1-q_{2}q_{3})(1-\mbox{\={r}}_{2}\mbox{\={r}}_{3})]
\label{pone}
\end{eqnarray} 
To derive \eqref{pone} we have added the contributions of four diagrams 
in Figure 1 and noted that
contributions of diagrams where a $Z$ line crosses a $ \phi$ line 
in the $Z$-$\phi$ interaction (the second and the fourth diagrams in Figure 1)
have a relative minus sign compared to the $Z$-$\phi$ interaction without crossing
(the first and the third diagrams in Figure 1).

Similarly from four diagrams of Figure 2 and from four diagrams of Figure 3 we get
\begin{eqnarray}
P_{2}=\sum_{0\leq l,0\leq k}^{l+k\leq {J-1}}(q_{2}\mbox{\={r}}_{2})^{l}(q_{3}
\mbox{\={r}}_{3})^{l+k}[(1-q_{2})(1-\mbox{\={r}}_{2})]q_{3}\mbox{\={r}}_{3}\\
P_{3}=\sum_{0\leq l,0\leq k}^{l+k\leq {J-1}}(q_{2}\mbox{\={r}}_{2})^{l}(q_{3}
\mbox{\={r}}_{3})^{l+k}[(1-q_{3})(1-\mbox{\={r}}_{3})]
\end{eqnarray}
\begin{figure}[h]
\begin{center}
\psfrag{phi1}{$\phi_{1}$}
\psfrag{phi2}{$\phi_{2}$}
\psfrag{phi3}{$\phi_{3}$}
\psfrag{x}{$x$}
\psfrag{a}{$a$}
\psfrag{b}{$b$}
\psfrag{c}{$c$}
\includegraphics[width=16cm]{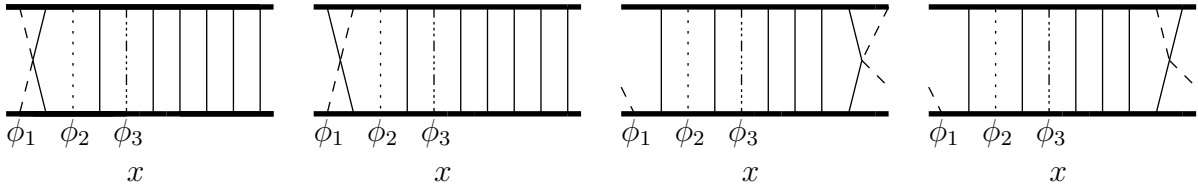}
\end{center}
\caption{Interacting diagrams for the 2-point 
function  where $ \phi_{1}$ interacts with $Z$.  
These diagrams give rise to $P_{1}$.}
\label{fig1}
\end{figure}

\begin{figure}[h]
\begin{center}
\psfrag{phi1}{$\phi_{1}$}
\psfrag{phi2}{$\phi_{2}$}
\psfrag{phi3}{$\phi_{3}$}
\psfrag{x}{$x$}
\psfrag{a}{$a$}
\psfrag{b}{$b$}
\psfrag{c}{$c$}
\includegraphics[width=16cm]{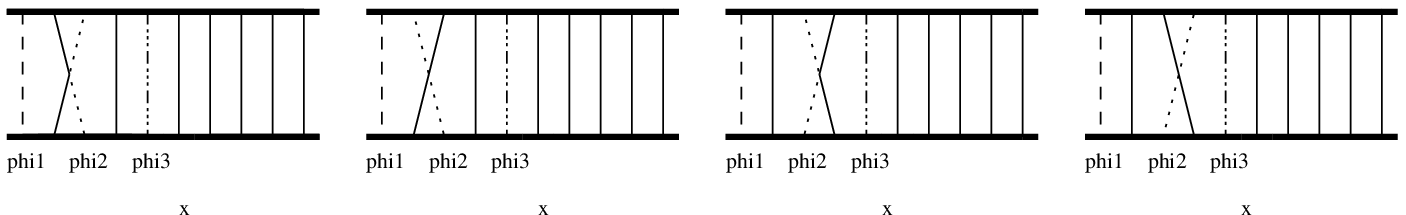}
\end{center}
\caption{Interacting diagrams for the 2-point function  
with $ \phi_{2}$ interactions. These diagrams give rise to $P_{2}$. }
\label{fig2}
\end{figure}

\begin{figure}[h]
\begin{center}
\psfrag{phi1}{$\phi_{1}$}
\psfrag{phi2}{$\phi_{2}$}
\psfrag{phi3}{$\phi_{3}$}
\psfrag{x}{$x$}
\psfrag{a}{$a$}
\psfrag{b}{$b$}
\psfrag{c}{$c$}
\includegraphics[width=16cm]{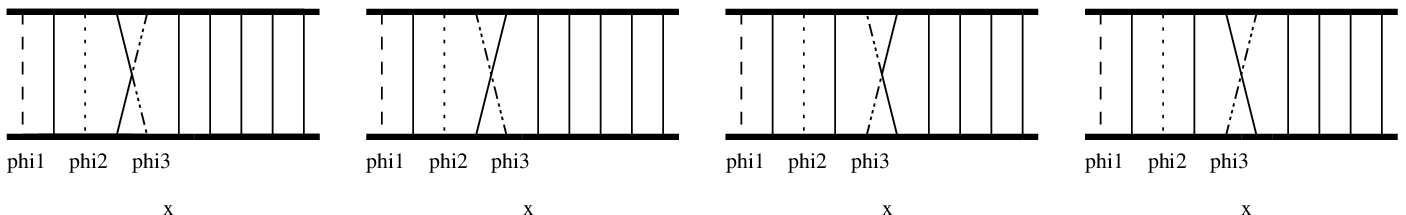}
\end{center}
\caption{Interacting diagrams for the 2-point function  
with $ \phi_{3}$ interactions. 
These diagrams give rise to $P_{3}$. }
\label{fig3}
\end{figure}

We now evaluate the double sum:
\begin{eqnarray*}
\sum_{0\leq l,0\leq k}^{l+k\leq {J-1}}(q_{2}
\mbox{\={r}}_{2})^{l}q_{3}\mbox{\={r}}_{3})^{l+k}=
\sum_{l=0}^{J-1}(q_{2}\mbox{\={r}}_{2}q_{3}
\mbox{\={r}}_{3})^{l}\sum_{k=0}^{J-1-l}(q_{3}\mbox{\={r}}_{3})^{k}
\end{eqnarray*}
\be
= \left\{ \begin{array}{ll}
            0        & \mbox{ when 
            $q_{2}\mbox{\={r}}_{2}\neq{1}$ and $q_{3}\mbox{\={r}}_{3}\neq{1}$}\\
            J(J+1)/2 & \mbox{ when 
            $q_{2}\mbox{\={r}}_{2}=1=q_{3}\mbox{\={r}}_{3}$}\\
            -\frac{J}{q_{2}\bar{r}_{2}-1} & \mbox{ when 
            $q_{2}\mbox{\={r}}_{2}\neq{1}$ and $q_{3}\mbox{\={r}}_{3}=1$}\\
             \frac{J}{q_{3}\bar{r}_{3}-1} & \mbox{ when 
             $q_{2}\mbox{\={r}}_{2}=1$ and $q_{3}\mbox{\={r}}_{3}\neq{1}$}
  \end{array} \right. 
  \label{arrr}
\ee  
It is clear that, in the BMN limit \eqref{doublel}, 
the correlator is non-zero only when 
$ q_{2}\mbox{\={r}}_{2}=1=q_{3}\mbox{\={r}}_{2}$, 
that is, when the operators in the correlator are the same.

The result for the second correlator,  
$\langle\bar{\cO}^{132}(0) \cO^{132}(x)\rangle$, 
is obtained from the first one by interchanging labels 2 and 3.
The sum of the two contributions,
$\langle\bar{\cO}^{123}(0) \cO^{123}(x)\rangle +
\langle\bar{\cO}^{132}(0) \cO^{132}(x)\rangle$, will have the second term
on the right hand side of \eqref{arrr} doubled up, and the third and fourth terms
cancelled.

We now show that the other two correlators
$\langle\bar{\cO}^{123}(0) \cO^{132}(x)\rangle$ and
$\langle\bar{\cO}^{132}(0) \cO^{123}(x)\rangle$,
vanish in our case. There are 12 diagrams to consider,
the first 6 are shown in Figure 4.            

\begin{figure}[h]
\begin{center}
\psfrag{phi1}{$\phi_{1}$}
\psfrag{phi2}{$\phi_{2}$}
\psfrag{phi3}{$\phi_{3}$}
\psfrag{x}{$x$}
\psfrag{a}{$a$}
\psfrag{b}{$b$}
\psfrag{c}{$c$}
\includegraphics[width=16cm]{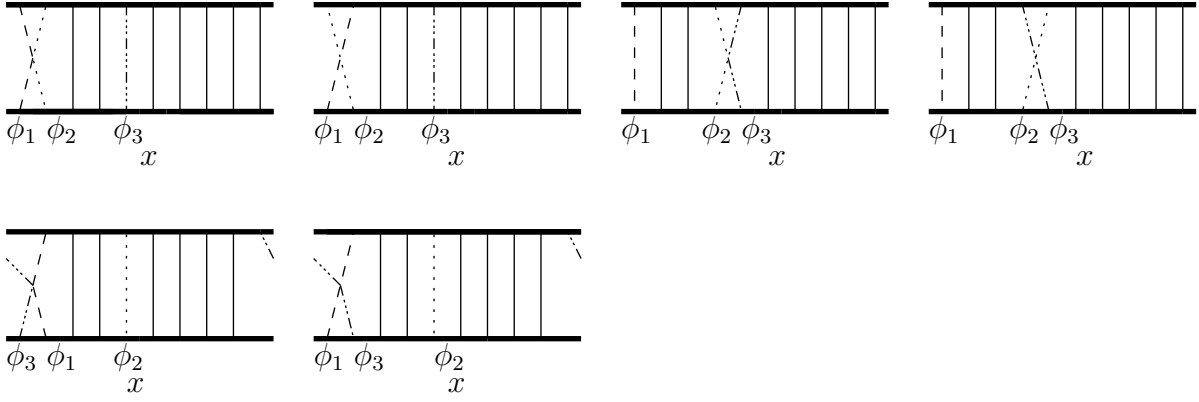}
\end{center}
\caption{Interacting diagrams for the two-point function.
These diagrams give rise to $P_{4}=0$. 
There are six additional diagrams with $\phi_{2}$ and  $\phi_{3}$ exchanged. 
Their sum is also zero. }
\label{fig4}
\end{figure}

The combined phase factor with these six diagrams is:
\begin{eqnarray}
P_{4}&=&\sum_{k=0}^{J}
(q_{2}^{0}q_{3}^{k}\mbox{\={r}}_{2}^{0}
\mbox{\={r}}_{3}^{k}-q_{2}^{0}q_{3}^{k}
\mbox{\={r}}_{2}^{J}\mbox{\={r}}_{3}^{k})\nonumber\\
&+& \sum_{l=0}^{J}(q_{2}^{l}q_{3}^{l}
\mbox{\={r}}_{2}^{l}\mbox{\={r}}_{3}^{l})
-q_{2}^{l}q_{3}^{l}\mbox{\={r}}_{2}^{l}
\mbox{\={r}}_{3}^{l})+\sum_{l=0}^{J}
(q_{2}^{l} q_{3}^{J}\mbox{\={r}}_{2}^{l}\mbox{\={r}}_{3}^{J}-q_{2}^{l} q_{3}^{0}\mbox{\={r}}_{2}^{l}\mbox{\={r}}_{3}^{J})\, =0 \ .
\end{eqnarray}
The remaining six diagrams are obtained from the ones in Figure 4
by exchanging $ \phi_{2}$ and  $ \phi_{3}$. 
They also sum to zero.
Thus, the non-diagonal terms do not contribute 
to the correlator at the planar level. 

Finally, 
combining with the free result we obtain
\begin{eqnarray}
&&\langle\bar{\cO}_{n_{1}n_{2}n_{3}}(0)
\cO_{\tilde{n}_{1}\tilde{n}_{2}\tilde{n}_{3}}(x)\rangle =
\delta_{{n}_{2}\tilde{n}_{2}}\delta_{n_{3}\tilde{n}_{3}}
\Delta(x)^{J+3}\, I(x) \\
&&\times \{1-g_{YM}^{2}N[(1- q_{2}q_{3})
(1-\bar{q}_{2}\bar{q}_{3})
+(1- q_{2})(1-\bar{q}_{2})+(1-\bar{q}_{3})(1-q_{3})]\},  \nonumber
\end{eqnarray}
where the four terms in curly brackets 
correspond respectively to the free contribution,
$P_1$, $P_2$ and $P_3$.
Now, substituting \eqref{Ix} for $I(x)$ with a subtraction \eqref{sub},
and using an expansion
\be
\Delta(x)^{\alpha} \simeq 1+\alpha\log \Delta(x)
=1+\alpha(-\log4\pi^{2}-\log x^{2})+O(\e))
\ee
we derive the final expression for the 2-point function,
\be \label{c2fin}
\langle\bar{\cO}_{n_{1}n_{2}n_{3}}(0)
\cO_{\tilde{n}_{1}\tilde{n}_{2}\tilde{n}_{3}}(x)\rangle
=\delta_{{n}_{2}\tilde{n}_{2}}
\delta_{n_{3},\tilde{n}_{3}}\Delta(x)^{J+3+\alpha}
[1-\alpha(\gamma+1-\log 4\pi-s)]   
\ee
where $\alpha$ denotes
\be
\alpha=\frac{\lambda'}{2}[(n_{2}+n_{3})^{2}+n_{2}^{2}+ n_{2}^{2}],
\ee 
and, from the right hand side of \eqref{c2fin}, it has to be identified with
the anomalous dimension of ${\cO}_{n_{1}n_{2}n_{3}}$,
\be
\Delta-J=\,3+\alpha=\, 3+ 
\frac{\lambda'}{2}[(n_{2}+n_{3})^{2}+n_{2}^{2}+ n_{2}^{2}], 
\ee
in agreement with dual string theory prediction.

The normalized operator is given by
\begin{eqnarray}
{\cO}_{n_{1}n_{2}n_{3}}
=\frac{1+\alpha/2(\gamma+1-\log 4\pi-s)}
{J \sqrt{N^{J+3}}}\sum_{0\leq l,k}^{l+k\leq {J}}
[q_{2}^{l}q_{3}^{l+k} \tr(\phi_{1}Z^{l}\phi_{2}Z^{k}
\phi_{3}Z^{J-l-k})\nonumber\\
+ q_{3}^{l}q_{2}^{l+k} 
\tr(\phi_{1}Z^{l}\phi_{3}Z^{k}\phi_{2}Z^{J-l-k})]
\label{3imnorm}
\end{eqnarray}

\section{Three-point functions}

Here our goal is evaluate 3-point functions involving BMN operators with 3 
scalar impurities in planar perturbation theory to order $\lambda'$.
We will consider two such 3-point functions,
\be \label{gdef}
 G_{3}(x_{1},x_{2})=
 \langle \bar{\cO}_{n_{1}n_{2}n_{3}}^{J}(0)
 \cO_{n_{1}' n_{2}' n_{3}'}^{J_{1}}(x_{1}) 
 \cO_{\rm vac}^{J_{2}}(x_{2})\rangle
\ee
and
\be \label{gpdef}
 G'_{3}(x_{1},x_{2})=
 \langle\bar{\cO}_{n_{1}n_{2}n_{3}}^{J}(0)
 \cO_{n_{2}' -n_{2}'}^{J_{1}}(x_{1}) \cO_{0}^{J_{2}}(x_{2})\rangle
\ee
Here ${\cO}_{n_{1}n_{2}n_{3}}^{J}$ 
with $n_{1}= -n_{2}-n_{3}$, is a $\Delta$-BMN operator with
3 scalar impurities, it is given by \eqref{3imnorm} plus multi-trace
expressions\footnote{In fact, it follows from the analysis of
\cite{BKPSS} that at the relevant to us first order in $g_2$,
only the double-trace corrections and only to
the barred operators in \eqref{gdef} and \eqref{gpdef} give    
non-vanishing contributions.}
at higher orders in $g_2$. The operator 
$\cO_{n_{2}' -n_{2}'}^{J_{1}}$ is a $\Delta$-BMN operator with
2 scalar impurities, at the classical single-trace level it is given by
\be
\cO_{n_{2}' -n_{2}'}^{J_{1}}=
\frac{1}{\sqrt{J N^{ J_{1}+2}}}
\sum_{l=0}^{ J_{1}}r_{2}^{l}\tr(\phi_{1}Z^{l}\phi_{2}Z^{J_1-l}),
\ee
and the remaining operators are chiral and are protected against 
quantum corrections,
\be
\cO_{0}^{J_{2}}=\frac{1}{\sqrt{N^{ J_{2}+1}}}\tr(\phi_{3}Z^{J_2}) \ , \quad
\cO_{\rm vac}^{J_{2}} =\frac{1}{\sqrt{J_2 N^{ J_{2}}}}\tr(Z^{J_2}).
\ee
We also note that the R-charge conservation implies that
\be
J_2 = J - J_1
\ee
In the first subsection we will derive conformal expressions \eqref{3pt}
and determine the 3-point coefficients $C_{123}$ for both of these Green functions
in the settings where the barred operator is general, and the two 
unbarred operators are chiral, i.e. correspond to two supergravity states.
As mentioned earlier, in this case, only the single-trace contributions
to the operators are relevant at the leading non-vanishing order in $g_2$,
thus simplifying our analysis significantly.

In the second subsection we will calculate the 3-point coefficients of
\eqref{gdef} and \eqref{gpdef} in the general case of two non-chiral operators.
Here the double-trace corrections are important, in order to derive the
conformal expression \eqref{3pt}. However, using a simple trick 
we will show how to
uniquely determine the coefficients $C_{123}$ directly from the single-trace
expressions for the operators, thus obtaining the main results of this Section,
Eqs. \eqref{cst123}, \eqref{c0123} and \eqref{ctil123}, \eqref{ctil0}.

\subsection{One string and two supergravity states}

As explained above, in this subsection only,
we set $n_{1}'= n_{2}'= n_{3}'=0$ and consider
\be \label{gdefch}
 G_{3}(x_{1},x_{2})=
 \langle \bar{\cO}_{n_{1}n_{2}n_{3}}^{J}(0)
 \cO_{000}^{J_{1}}(x_{1}) 
 \cO_{\rm vac}^{J_{2}}(x_{2})\rangle
\ee
and
\be \label{gpdefch}
 G'_{3}(x_{1},x_{2})=
 \langle\bar{\cO}_{n_{1}n_{2}n_{3}}^{J}(0)
 \cO_{00}^{J_{1}}(x_{1}) \cO_{0}^{J_{2}}(x_{2})\rangle
\ee
At first we consider the 3-point function $G_{3}(x_{1},x_{2})$ 
and express it as follows: 
\begin{eqnarray} 
G_{3}(x_{1},x_{2})=
\frac{1+\frac{\alpha}{2}(\gamma+1-\log 4\pi-s)}
{J \sqrt{N^{J+3}}}
\frac{N^{J+2}J_{2}}{J_{1}\sqrt{J_{2}N^{J_{1}+3}N^{J_{2}}}}
\nonumber\\
\Delta(x_{1})^{J_{1}+3}\Delta(x_{2})^{J_{2}}\,
(X-\lambda Y K(x_{1},x_{2}))
\label{g3first}
\end{eqnarray}
where $\lambda=g^2_{YM}N$ and $X$ and $Y$ are the combined phase-factors 
at the free and interacting level respectively.
$ K(x_{1},x_{2})$ is the interaction integral for the diagrams 
depicted on Figures 6--9 (as in \cite{CKT}):
\begin{eqnarray}
 K(x_{1},x_{2})=\left(\frac{\Gamma(1-\e)}{4 \pi^{2-\e}} \right)^{2}
 (x_{1}^{2})^{1-\e}(x_{2}^{2})^{1-\e}\int 
 \frac{ d^{4-2\e}y}{(y^{2})^{2-2\e}(y-x_{1})^{2(1-\e)}(y-x_{2})^{2(1-\e)}}
 \nonumber\\
=\frac{1}{16\pi^{2}}
(\frac{1}{\e}+\gamma+2+\log\pi+
\log\frac{x_{1}^{2}x_{2}^{2}}{x_{12}^{2}}+\emph{O}(\e))
\end{eqnarray}
The first fraction on the right hand side of \eqref{g3first}
arises from the normalization of $ \bar{\cO}_{n_{1}n_{2}n_{3}}^{J}$. 
The denominator of the second fraction arises from the normalizations 
of the other two operators, while the summation over the $J+2$ loops 
gives a factor of $N^{J+2}$. 
The remaining factor of $J_{2}$
comes from the Wick contractions with $ \cO_{\rm vac}^{J_{2}}$.

Free diagrams are shown in Figure 5. 
There are six diagrams because of the six different ways 
of arranging three  $ \phi$'s in a trace. 
We denote with $X_{123}$ the combined phase factor of a free diagram where
$ \phi_{1}$ comes first, $ \phi_{2}$ is second and  $ \phi_{3}$ is third. 

\begin{figure}[h]
\begin{center}
\psfrag{phi1}{$\phi_{1}$}
\psfrag{phi2}{$\phi_{2}$}
\psfrag{phi3}{$\phi_{3}$}
\psfrag{x1}{$x_{1}$}
\psfrag{x2}{$x_{2}$}
\psfrag{a}{$a$}
\psfrag{b}{$b$}
\psfrag{c}{$c$}
\includegraphics[height=4cm]{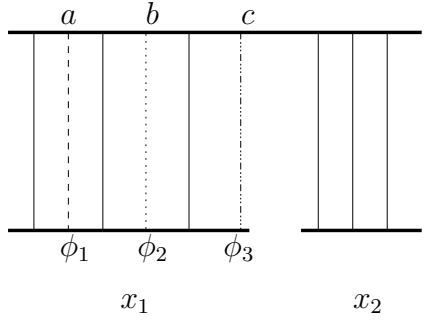}
\end{center}
\caption{A typical free diagram for $G_{3}$. 
This  diagram gives rise to $X_{123}$. There are five additional diagrams 
with $\phi_{1}$, $\phi_{2}$ and $\phi_{3}$ interchanged.
$a$, $b$ and $c$ are the positions of the impurities in the trace. For
this particular diagram, $a=2$, $b=4$, $c=6$.}
\label{fig5}
\end{figure} 

We have:
\begin{eqnarray}
X_{123}=&&\sum_{a=1}^{J_{1}+1}\sum_{b=a+1}^{J_{1}+2}
\sum_{c=b+1}^{J_{1}+3}\mbox{\={q}}_{2}^{b-a-1}\mbox{\={q}}_{3}^{c-a-2} 
\stackrel{J\to \infty}{\longrightarrow}\nonumber\\
&&J^{3}\int_{0}^{J_{1}/J}da\int_{a}^{J_{1}/J}db\int_{b}^{J_{1}/J}dc \, 
e^{-2\pi\imath n_{2}(b-a)} e^{-2\pi\imath n_{3}(c-a)}
\end{eqnarray}
\begin{eqnarray}
X_{231}=&&\sum_{a=1}^{J_{1}+1}\sum_{b=a+1}^{J_{1}+2}
\sum_{c=b+1}^{J_{1}+3}\mbox{\={q}}_{2}^{J-(c-a-2)}\mbox{\={q}}_{3}^{J-(c-b-1)}
\stackrel{J\to \infty}{\longrightarrow}\nonumber\\
&&J^{3}\int_{0}^{J_{1}/J}da\int_{a}^{J_{1}/J}db
\int_{c}^{J_{1}/J}dc \,
e^{-2\pi\imath n_{2}(-1)(c-a)} e^{-2\pi\imath n_{3}(-1)(c-b)}
\end{eqnarray}
\begin{eqnarray}
X_{312}=&&\sum_{a=1}^{J_{1}+1}\sum_{b=a+1}^{J_{1}+2}
\sum_{c=b+1}^{J_{1}+3}\mbox{\={q}}_{2}^{c-b-1}
\mbox{\={q}}_{3}^{J-(b-a-1)}\stackrel{J\to \infty}{\longrightarrow}
\nonumber\\
&&J^{3}\int_{0}^{J_{1}/J}da
\int_{a}^{J_{1}/J}db\int_{b}^{J_{1}/J}dc \, 
e^{-2\pi i n_{2}(c-b)} e^{-2\pi i n_{3}(-1)(b-a)}\\
\end{eqnarray}
where $ a,b,c$ are the positions of the first, second and third impurities 
in the trace. Also it is easy to see that $X_{132}$ is equal 
to $X_{123}$ with  $\mbox{\={q}}_{2}$ 
and $\mbox{ \={q}}_{3}$ exchanged, $X_{213}$ is $X_{312}$ 
with $\mbox{\={q}}_{2}$ and $\mbox{\={q}}_{3}$ exchanged and, 
$X_{321}$ is $X_{231}$ with $\mbox{\={q}}_{2}$ and 
$\mbox{\={q}}_{3}$ exchanged. 

The sum of the six $X$'s is:
\be
\label{twentyone}
X=J^{3}\int_{0}^{J_{1}/J}da\int_{0}^{J_{1}/J}db
\int_{0}^{J_{1}/J}dc \, e^{-2\pi\imath n_{2}(b-a)} 
e^{-2\pi\imath n_{3}(c-a)}
\ee
For example the part of the above sum with $c>b>a$ is $X_{123}$, 
the part with $c>a>b$ is $X_{213}$ and so on. 
Evaluating the integral we get
\be
X=J^{3}2^{3} \, \frac{\sin(\pi n_{2}J_{1}/J)\sin(\pi n_{3}J_{1}/J)
\sin(\pi (n_{2}+n_{3})J_{1}/J)}{(2\pi)^{3}
(n_{2}+n_{3})n_{2}n_{3}}
\ee

We now calculate the phase factors coming from interacting planar 
diagrams.
In the case where  $\phi_{1}$ interacts with $Z$ we have eight diagrams with the
first four shown in Figure 6 and the remaining four obtained by
interchanging $\phi_{2}$ and  $\phi_{3}$. We do not need to consider diagrams where
$\phi_i$ interacts with $\phi_j$ since they will be suppressed in the BMN limit
\eqref{doublel} relative to $\phi$-$Z$ interactions of Figure 6. 
 
\begin{figure}[h]
\begin{center}
\psfrag{phi1}{$\phi_{1}$}
\psfrag{phi2}{$\phi_{2}$}
\psfrag{phi3}{$\phi_{3}$}
\psfrag{x1}{$x_{1}$}
\psfrag{x2}{$x_{2}$}
\psfrag{a}{$a$}
\psfrag{b}{$b$}
\psfrag{c}{$c$}
\includegraphics[width=16cm]{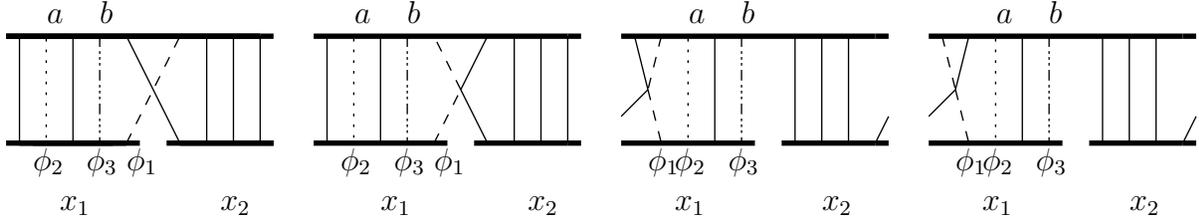}
\end{center}
\caption{Interacting diagrams for $G_{3}$. 
These diagrams give rise to $Y_{123}$. 
There are four additional diagrams with    
$\phi_{2}$ and  $\phi_{3}$ exchanged.}
\label{fig6}
\end{figure} 

The phase combined factor of the four diagrams in Figure 6 is:
\begin{eqnarray}
Y_{123}=\sum_{a=1}^{J_{1}+1}\sum_{b=a+1}^{J_{1}+2}
\left[-\mbox{\={q}}_{2}^{J-(J_{1}+3-a-1)}\mbox{\={q}}_{3}^{J-(J_{1}+3-b)}
+\mbox{\={q}}_{2}^{J-(J_{1}+3-a-2)}\mbox{\={q}}_{3}^{J-(J_{1}+3-b-1)}\right]
\nonumber\\
+\sum_{a=3}^{J_{1}+2}\sum_{b=a+1}^{J_{1}+3}\left[\mbox{\={q}}_{2}^{a-3}
\mbox{\={q}}_{3}^{b-4}-\mbox{\={q}}_{2}^{a-2}\mbox{\={q}}_{3}^{b-3}\right]
\nonumber\\
=\sum_{a=1}^{J_{1}+1}\sum_{b=a+1}^{J_{1}+2}\mbox{\={q}}_{2}^{-(J_{1}-a+1)}
\mbox{\={q}}_{3}^{-(J_{1}-b+2)}(1-\mbox{\={q}}_{2}^{-1}
\mbox{\={q}}_{3}^{-1})+\sum_{a=3}^{J_{1}+2}\sum_{b=a+1}^{J_{1}
+3}\mbox{\={q}}_{2}^{a-3}\mbox{\={q}}_{3}^{b-4}
(1-\mbox{\={q}}_{2}\mbox{\={q}}_{3})
\nonumber\\
\rightarrow\frac{2\pi i (n_{2}+n_{3})}{J}\left[\sum_{a=3}^{J_{1}+2}
\sum_{b=a+1}^{J_{1}+3}
\mbox{\={q}}_{2}^{a-3}\mbox{\={q}}_{3}^{b-4}
-\sum_{a=1}^{J_{1}+1}\sum_{b=a+1}^{J_{1}+2}\mbox{\={q}}_{2}^{a-1}
\mbox{\={q}}_{3}^{b-2}\mbox{\={q}}_{2}^{-J_{1}}
\mbox{\={q}}_{3}^{-J_{1}}\right]
\end{eqnarray}
In deriving the last line above, we have used that 
$1-\mbox{\={q}}_{2}\mbox{\={q}}_{3}
\rightarrow \frac{2\pi i (n_{2}+n_{3})}{J}$ in the BMN limit. 
Converting the sum into an integral we get
\begin{eqnarray} 
Y_{123}={2\pi i (n_{2}+n_{3})}{J}
\int_{0}^{J_{1}/J}da\int_{a}^{J_{1}/J}db 
\, e^{-2\pi i n_{2}a} e^{-2\pi i n_{3}b}
[1- e^{-2\pi i (n_{3}+n_{2})(-J_{1}/J)}]
\nonumber\\
=
-J [1- e^{-(A_2+A_3)J_{1}/J}]
\frac{e^{(A_{3}+A_{2})J_{1}/J}A_{3}
-e^{A_{3}J_{1}/J}(A_{2}+A_{3})
+A_{2}}{A_{3}A_{2}}
\label{fourtyfour}
\end{eqnarray}
where $A_{2}=-2\pi i n_{2}$ and $A_{3}=-2\pi i n_{3}$. 
We must add 4 more diagrams with $ \phi_{2}$ and $ \phi_{3}$ 
exchanged. This gives the expression above with  
$n_{2}$ and $n_{3}$  exchanged. If we sum the two contributions we get:
\begin{eqnarray}
Y_{123}+Y_{132}=\frac{4(n_{3}+n_{2})^{2}J
\sin(\pi n_{2}J_{1}/J)\sin(\pi n_{3}J_{1}/J)
\sin(\pi(n_{3}+n_{2}) J_{1}/J)}{\pi(n_{3}+n_{2})n_{2}n_{3}}
\end{eqnarray}

In the case where  $\phi_{2}$ interacts we have again eight diagrams, 
see Figure 7. 
\begin{figure}[h]
\begin{center}
\psfrag{phi1}{$\phi_{1}$}
\psfrag{phi2}{$\phi_{2}$}
\psfrag{phi3}{$\phi_{3}$}
\psfrag{x1}{$x_{1}$}
\psfrag{x2}{$x_{2}$}
\psfrag{a}{$a$}
\psfrag{b}{$b$}
\psfrag{c}{$c$}
\includegraphics[width=16cm]{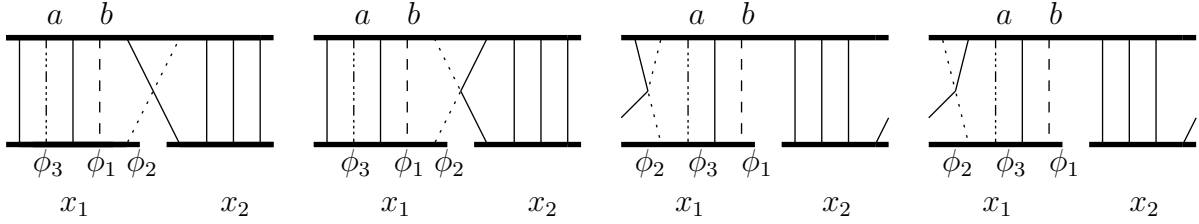}
\end{center}
\caption{Interacting diagrams for $G_{3}$. 
These diagrams give rise to $Y_{231}$. 
There are four additional diagrams with $\phi_{1}$ and $\phi_{3}$ exchanged.}
\label{fig7}
\end{figure}

The combined phase factor of four diagrams in Figure 7 is easily
obtained by substituting $n_2\to n_3$ and $n_3\to-(n_2+n_3)$ 
into \eqref{fourtyfour}, as follows from
comparing diagrams in Figures 6 and 7 and remembering that $n_1:=-(n_2+n_3)$.
In the BMN limit we have,
\be \label{y231}
Y_{231}
=J [1- e^{A_2J_{1}/J}]
\frac{-e^{-A_{2}J_{1}/J}(A_{2}+A_{3})
+e^{-(A_{3}+A_{2})J_{1}/J}A_{2}
+A_{3}}{A_{3}(A_{2}+A_{3})}
\ee
Now we consider the above diagrams with  $ \phi_{1}$ and $ \phi_{3}$ 
exchanged, i.e. $-(n_2+n_3) \leftrightarrow n_3$ and $n_2$ unchanged in
\eqref{y231}, 
\be
Y_{213}=J [1- e^{A_2J_{1}/J}]
\frac{e^{-A_{2}J_{1}/J}A_3
+e^{A_3 J_{1}/J}A_{2}-A_2-A_{3}}{A_{3}(A_{2}+A_{3})}
\ee
Summing the above contributions we arrive at
\begin{eqnarray}
Y_{231}+Y_{213}=\frac{4n_{2}^{2}J\sin(\pi n_{2}J_{1}/J)
\sin(\pi n_{3}J_{1}/J)\sin(\pi(n_{3}+n_{2}) 
J_{1}/J)}{\pi(n_{3}+n_{2})n_{2}n_{3}} \ .
\end{eqnarray}

Similarly, for diagrams in Figure 8 and their four partners,
 we find:
\begin{eqnarray}
Y_{312}+Y_{321}=\frac{4n_{3}^{2}J\sin(\pi n_{2}J_{1}/J)
\sin(\pi n_{3}J_{1}/J)\sin(\pi(n_{3}+n_{2}) J_{1}/J)}
{\pi(n_{3}+n_{2})n_{2}n_{3}}
\end{eqnarray}
\begin{figure}[h]
\begin{center}
\psfrag{phi1}{$\phi_{1}$}
\psfrag{phi2}{$\phi_{2}$}
\psfrag{phi3}{$\phi_{3}$}
\psfrag{x1}{$x_{1}$}
\psfrag{x2}{$x_{2}$}
\psfrag{a}{$a$}
\psfrag{b}{$b$}
\psfrag{c}{$c$}
\includegraphics[width=16cm]{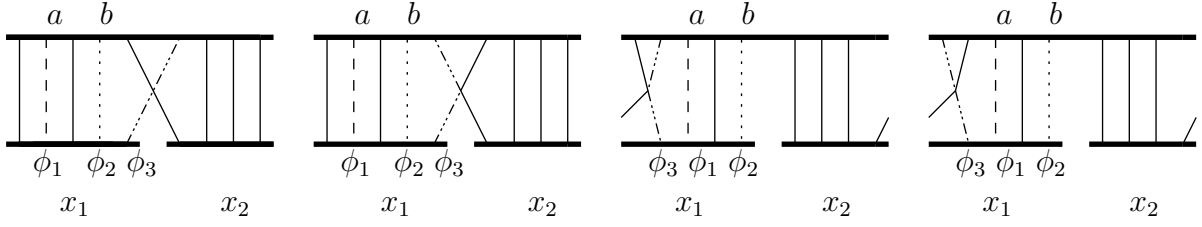}
\end{center}
\caption{Interacting diagrams for $G_{3}$, contributing to $Y_{312}$. 
There are four additional diagrams with $\phi_{1}$ and $\phi_{2}$ exchanged.}
\label{fig8}
\end{figure}

For completeness, we note that diagrams without $x_1$-to-$x_2$ connection,
depicted in Figure 9, sum to zero.
\begin{figure}[h]
\begin{center}
\psfrag{9a}{$9a$}
\psfrag{9b}{$9b$}
\psfrag{9c}{$9c$}
\psfrag{9d}{$9d$}
\psfrag{9e}{$9e$}
\psfrag{9f}{$9f$}
\psfrag{9g}{$9g$}
\psfrag{9h}{$9h$}
\psfrag{phi1}{$\phi_{1}$}
\psfrag{phi2}{$\phi_{2}$}
\psfrag{phi3}{$\phi_{3}$}
\psfrag{x1}{$x_{1}$}
\psfrag{x2}{$x_{2}$}
\psfrag{a}{$a$}
\psfrag{b}{$b$}
\psfrag{c}{$c$}
\includegraphics[width=16cm]{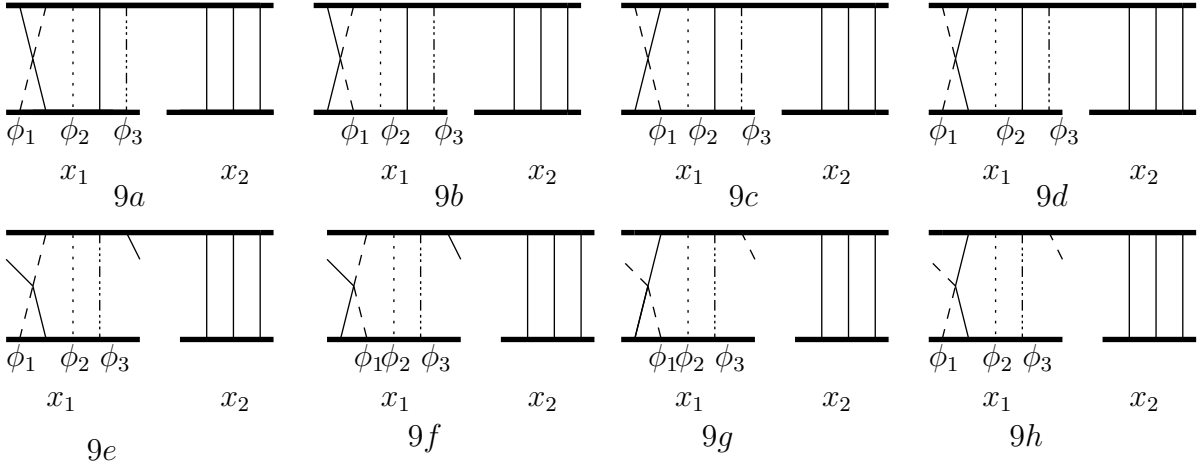}
\end{center}
\caption{Interacting diagrams for $G_{3}$. 
Diagrams $ 9a, 9c, 9e$ and $9g$ have an opposite sign with respect 
to $9b, 9d, 9f$ and $9h$, so they cancel pairwise. 
There are  additional diagrams where $\phi_{2}$ and  $\phi_{3}$ are 
exchanged which also add up zero. 
Additional diagrams where $\phi_{2}$ or $\phi_{3}$ (rather than $\phi_1$
interact with $Z$ also cancel in the same way.}
\label{fig9}
\end{figure}

Putting together all the expressions above, we get for $G_3$
\begin{eqnarray}
 & & G_{3}(x_{1},x_{2})=\frac{1+\alpha/2(\gamma+1-\log 4\pi-s)}
 {J \sqrt{N^{J+3}}}\frac{N^{J+2}J_{2}}{J_{1}
\sqrt{J_{2}N^{J_{1}+3}N^{J_{2}}}}\, \Delta(x_{1})^{J_{1}+3}
\Delta(x_{2})^{J_{2}}
\nonumber\\
& & \qquad \times \frac{J^{3}}{\pi^{3}}\frac{\sin(\pi n_{2}J_{1}/J)
\sin(\pi n_{3}J_{1}/J)
\sin(\pi(n_{3}+n_{2}) J_{1}/J)}{(n_{3}+n_{2})n_{2}n_{3}}\nonumber\\
 & &\times \left[1-\frac{\l'}{4}((n_{3}+n_{2})^{2}+n_{2}^{2}
+n_{3}^{2}))(\frac{1}{\e}+\gamma+2
+\log\pi+\log\frac{x_{1}^{2} x_{2}^{2}}{x_{12}^{2}})\right]
\end{eqnarray}
Subtracting $1/\e+s$ we obtain the result
\be \label{g3res1}
G_{3}(x_{1},x_{2})=C_{123}\ 
\Delta(x_{1})^{J_{1}+3+
 \alpha/2}\Delta(x_{2})^{J_{2}+\alpha/2}\Delta(x_{12})^{-\alpha/2}
\ee
with
\be \label{cg3res1}
C_{123}=\frac{J^{2}\sqrt{J_{2}}}{J_{1}N \pi^{3}}\, 
\frac{\sin(\pi n_{2}J_{1}/J)
 \sin(\pi n_{3}J_{1}/J)\sin(\pi(n_{3}+n_{2}) 
 J_{1}/J)}{(n_{3}+n_{2})n_{2}n_{3}}
 \, \left(1-\frac{\alpha}{2}\right)
\ee
A few comments are in order.
First, we note that we have proved that to order 
in $\l'$ and $g_2$ we are working here,
$G_{3}$ takes
the conformal form of \eqref{3pt}. This is so since
\eqref{g3res1} is nothing other than the conformal
expression \eqref{3pt} for $G_3$. Second, we have derived 
the expression for the coefficient $C_{123}$ given by
\eqref{cg3res1}. This expression does not depend on $s$
and, hence, is the subtraction scheme independent, as expected.
In what follows, and in parallel with \cite{CK,BKPSS}, we will
use only the leading order in $\l'$ part of $C_{123}$, i.e.
will set $\alpha=0$ in \eqref{cg3res1}. This is because
the, yet unaccounted, mixing effects at order $\l'$ can change
{\it constant} order $\l'$ contributions to  $C_{123}$
(but not the logarithms in eqref{g3res1}, which cannot appear in
the $x$-independent mixing matrices).

 \medskip
 
 {\it Three-point correlator $ G'_{3}(x_{1},x_{2})$}

We now consider the second 3-point function, $ G'_{3}(x_{1},x_{2})$,
of Eq. \eqref{gpdefch}.
Its structure is much the same as for $ G_{3}(x_{1},x_{2})$ 
leading to the following expression
\begin{eqnarray}
G'_{3}(x_{1},x_{2})=\frac{1+\alpha/2(\gamma+1-\log 4\pi-s)}
{J \sqrt{N^{J+3}}}\frac{N^{J+2}}{\sqrt{J_{1}N^{J_{1}+2}N^{J_{2}+1}}}
\nonumber\\
\times \Delta(x_{1})^{J_{1}+2}\Delta(x_{2})^{J_{2}+1}
(P-\lambda Q K(x_{1},x_{2}))
\end{eqnarray}
where $P$ and $Q$ are the phase factors to be determined shortly.

$P$ is the phase factor which we get by
summing the contributions from the two free diagrams depicted in 
Figure 10.
\begin{figure}[h]
\begin{center}
\psfrag{phi1}{$\phi_{1}$}
\psfrag{phi2}{$\phi_{2}$}
\psfrag{phi3}{$\phi_{3}$}
\psfrag{x1}{$x_{1}$}
\psfrag{x2}{$x_{2}$}
\psfrag{a}{$a$}
\psfrag{b}{$b$}

\psfrag{c}{$c$}
\includegraphics[width=10cm,height=4cm]{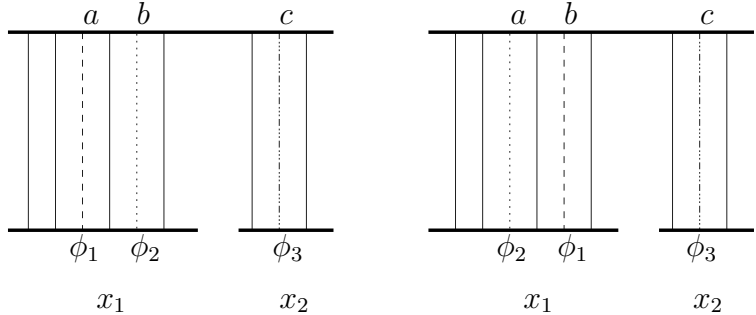}
\end{center}
\caption{ Free diagrams for $G_{3}$. This  diagram gives rise to $P$.}
\label{fig10}
\end{figure} 

\begin{eqnarray}
P=\sum_{a=1}^{J_{1}+1}\sum_{b=a+1}^{J_{1}+2}
\sum_{c=J_{1}+3}^{J+3}\mbox{\={q}}_{2}^{b-a-1}
\mbox{\={q}}_{3}^{c-a-2}+\sum_{a=1}^{J_{1}+1}
\sum_{b=a+1}^{J_{1}+2}\sum_{c=J_{1}+3}^{J+3}
\mbox{\={q}}_{2}^{J-(b-a-1)}\mbox{\={q}}_{3}^{c-b-1}
\end{eqnarray}
Converting the above sun into an integral one finds
 \begin{eqnarray}
P=J^{3}\,\int_{0}^{J_{1}/J}da\int_{a}^{J_{1}/J}db
\int_{J_{1}/J}^{1}dc \, e^{-2\pi i n_{2}(b-a)} 
e^{-2\pi i n_{3}(c-a)}
\nonumber\\
+J^{3}\,\int_{0}^{J_{1}/J}da\int_{a}^{J_{1}/J}db\int_{J_{1}/J}^{1}dc
\, e^{-2\pi i n_{2}(-1)(b-a)} e^{-2\pi i n_{3}(c-b)}
\nonumber\\
=J^{3}\int_{0}^{J_{1}/J}da
\int_{0}^{J_{1}/J}db\int_{J_{1}/J}^{1}dc 
\, e^{-2\pi i n_{2}(b-a)} e^{-2\pi i n_{3}(c-a)}
\end{eqnarray}
Evaluating the integral we finally arrive at
\begin{eqnarray}
P=-J^{3}\frac{\sin(\pi n_{2}J_{1}/J)
\sin(\pi n_{3}J_{1}/J)\sin(\pi(n_{3}+n_{2}) J_{1}/J)}
{\pi^{3}(n_{3}+n_{2})n_{2}n_{3}} \ .
\end{eqnarray}

Now we have to account for the interacting diagrams. 
In the case where $\phi_{1}$ takes part in the interaction 
we have four diagrams which are shown in Figure 11.
The corresponding phase factor is
\begin{eqnarray}
Q_{1}=\sum_{a=1}^{J_{1}+1}\sum_{b=J_{1}+4}^{J+3}-
\mbox{\={q}}_{2}^{J-(J_{1}+3-a)}\mbox{\={q}}_{3}^{b-(J_{1}+3)}
+\mbox{\={q}}_{2}^{J-(J_{1}+2-a)}\mbox{\={q}}_{3}^{b-(J_{1}+2)}
\nonumber\\
+\sum_{a=3}^{J_{1}+2}\sum_{b=J_{1}+3}^{J+2}
\mbox{\={q}}_{2}^{a-3}\mbox{\={q}}_{3}^{b-4}
-\mbox{\={q}}_{2}^{a-2}\mbox{\={q}}_{3}^{b-3} \ ,  
\end{eqnarray}
which in the BMN limit is
\begin{eqnarray}
Q_{1}={2\pi i( n_{2}+ n_{3})}J
(1- e^{-2\pi i(n_{3}+n_{2})(-1)J_{1}/J})
\int_{0}^{J_{1}/J}da
\int_{J_{1}/J}^{1}db e^{-2\pi i n_{2}a} 
e^{-2\pi i n_{3}b}\nonumber\\
=-\frac{4}{\pi}J(n_{3}+n_{2})^{2}
\frac{\sin(\pi n_{2}J_{1}/J)
\sin(\pi n_{3}J_{1}/J)
\sin(\pi(n_{3}+n_{2}) J_{1}/J)}{(n_{3}+n_{2})n_{2}n_{3}}
\end{eqnarray}
\begin{figure}[h]
\begin{center}
\psfrag{phi1}{$\phi_{1}$}
\psfrag{phi2}{$\phi_{2}$}
\psfrag{phi3}{$\phi_{3}$}
\psfrag{x1}{$x_{1}$}
\psfrag{x2}{$x_{2}$}
\psfrag{a}{$a$}
\psfrag{b}{$b$}
\psfrag{c}{$c$}
\includegraphics[width=16cm]{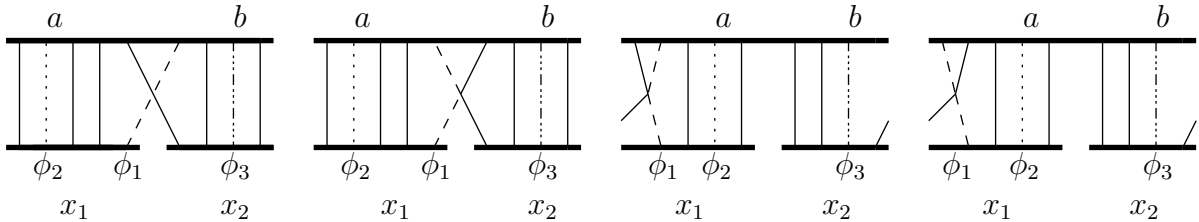}
\end{center}
\caption{Interacting diagrams for $G'_{3}$ 
contributing to $Q_1$. The $\phi_{1}$ line is in the $a^{th}$ 
position in the $\bar{O}_{n_{2},n_{3}}^{J}(0)$ trace and
$\phi_{3}$ is in the $b^{th}$ position.}
\label{fig11}
\end{figure}

In the case where $\phi_{2}$ takes part in the interaction 
we have again four diagrams which are shown in Figure 12.
The corresponding phase factor is
\begin{eqnarray}
Q_{2}=\sum_{a=1}^{J_{1}+1}\sum_{b=J_{1}
+4}^{J+3}-\mbox{\={q}}_{2}^{J_{1}+2-a)}
\mbox{\={q}}_{3}^{b-a-2}+\mbox{\={q}}_{2}^{J_{1}+-a)}
\mbox{\={q}}_{3}^{b-a-2}
\nonumber\\
+\sum_{a=3}^{J_{1}+2}\sum_{b=J_{1}+3}^{J+2}
\mbox{\={q}}_{2}^{-(a-3)}\mbox{\={q}}_{3}^{b-a-1}
-\mbox{\={q}}_{2}^{-(a-2)}\mbox{\={q}}_{3}^{b-a-1}\ ,
\end{eqnarray}
which in the BMN limit becomes
\begin{eqnarray}
Q_{2}=-J^{2}\frac{2\pi i n_{2}}{J}(1- e^{-2\pi i n_{2}J_{1}/J})
\int_{0}^{J_{1}/J}da\int_{J_{1}/J}^{1}db \,
e^{-2\pi i n_{2}(-a)} e^{-2\pi i n_{3}(b-a)}
\nonumber\\
=-\frac{4}{\pi}J n_{2}^{2}\,
\frac{\sin(\pi n_{2}J_{1}/J)\sin(\pi n_{3}J_{1}/J)
\sin(\pi(n_{3}+n_{2}) J_{1}/J)}{(n_{3}+n_{2})n_{2}n_{3}}
\end{eqnarray}
\begin{figure}[h]
\begin{center}
\psfrag{phi1}{$\phi_{1}$}
\psfrag{phi2}{$\phi_{2}$}
\psfrag{phi3}{$\phi_{3}$}
\psfrag{x1}{$x_{1}$}
\psfrag{x2}{$x_{2}$}
\psfrag{a}{$a$}
\psfrag{b}{$b$}
\psfrag{c}{$c$}
\includegraphics[width=16cm]{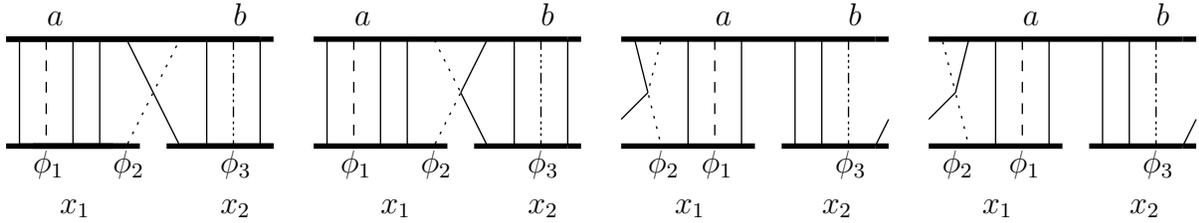}
\end{center}
\caption{Interacting diagrams for $G'_{3}$ contributing to $Q_2$.}
\label{fig12}
\end{figure}

For interacting $\phi_{3}$ there are eight diagrams. 
The first four are depicted in Figure 13, and the other four 
are obtained by exchanging $\phi_{1}$ with $\phi_{2}$. 
The phase factor associated with the four diagrams of Figure 13
in the BMN limit is
\begin{eqnarray}
Q_{3}^{(1)}=J^{2}\frac{2\pi i n_{3}}{J}
(1- e^{-2\pi i n_{3}J_{1}/J})\int_{0}^{J_{1}/J}da
\int_{a}^{J_{1}/J}db \, e^{-2\pi i n_{2}(b-a)} 
e^{-2\pi i n_{3}(-a)}
\end{eqnarray}
The phase factor $Q_{3}^{(2)}$ for the remaining four diagrams is obtained
by exchanging $n_1 \leftrightarrow n_2$ in $Q_{3}^{(1)}$. 
Adding the two we obtain
\begin{eqnarray}
Q_{3}=Q_{3}^{(1)}+Q_{3}^{(2)}=
{2\pi i n_{3}}J
(1- e^{-2\pi i n_{3}J_{1}/J})\int_{0}^{J_{1}/J}da
\int_{0}^{J_{1}/J}db \,
e^{-2\pi i n_{2}(b-a)} e^{-2\pi i n_{3}(-a)}
\nonumber\\
=-\frac{4}{\pi}J n_{3}^{2}\,
\frac{\sin(\pi n_{2}J_{1}/J)
\sin(\pi n_{3}J_{1}/J)\sin(\pi(n_{3}
+n_{2}) J_{1}/J)}{(n_{3}+n_{2})n_{2}n_{3}}
\end{eqnarray}
\begin{figure}[ht]
\begin{center}
\psfrag{phi1}{$\phi_{1}$}
\psfrag{phi2}{$\phi_{2}$}
\psfrag{phi3}{$\phi_{3}$}
\psfrag{x1}{$x_{1}$}
\psfrag{x2}{$x_{2}$}
\psfrag{a}{$a$}
\psfrag{b}{$b$}
\psfrag{c}{$c$}
\includegraphics[width=16cm]{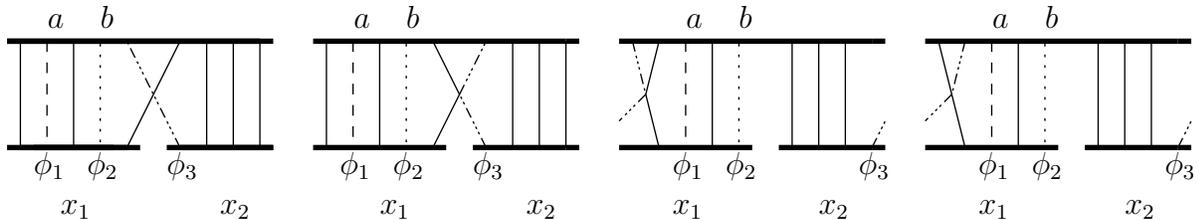}
\end{center}
\caption{Interacting diagrams for $G'_{3}$ contributing to $ Q_3$. 
There are also diagrams with $\phi_{1}$ 
and $\phi_{2}$ exchanged.}
\label{fig13}
\end{figure}

Taking everything into account, our final expression for
$G'_{3}$ takes the conformal form
\begin{eqnarray}
 G'_{3}(x_{1},x_{2})= C'_{123}\Delta(x_{1})^{J_{1}+2+\alpha/2}
\Delta(x_{2})^{J_{2}+1+\alpha/2}\Delta(x_{12})^{-\alpha/2}
\end{eqnarray}
with the 3-point coefficient
\begin{eqnarray}
\label{cprch}
 C'_{123}=-\frac{J^{2}}{\sqrt{J_{1}}N\pi^{3}} 
 \frac{\sin(\pi n_{2}J_{1}/J)
 \sin(\pi n_{3}J_{1}/J)\sin(\pi(n_{3}
 +n_{2}) J_{1}/J)}{(n_{3}+n_{2})n_{2}n_{3}}\, (1-\alpha/2)
\end{eqnarray}
This expression is again subtraction scheme independent.
As in the case of $G_3$ discussed earlier, we will set 
$\alpha=0$ on the right hand side of \eqref{cprch} to be safe from
unknown mixing effects at order $\l'$.

\subsection{Three-point functions with two string states}

We are now ready to finally address the general case and
calculate the 3-point coefficients of
\eqref{gdef} and \eqref{gpdef} for two non-chiral operators.
Here the mixing of the known single-trace BMN operators
with double-trace corrections is important as it does contribute
to the conformal expression \eqref{3pt}. 
However, our goal is not to derive the conformal expression
on the right hand side of  \eqref{3pt} (which must be correct anyway,
as far as the mixing effects are such that we are dealing
with scalar conformal primary operators). Our goal is to calculate
the coefficient $C_{123}$. At leading order, the only mixing effect which 
contributes to the right hand side of  \eqref{3pt} is the mixing with 
the double-traces of the barred operator 
$\bar{\cO}_{n_{1}n_{2}n_{3}}^{J}(0)$
in \eqref{gdef} and \eqref{gpdef}. These mixing effects will affect
the free-theory contribution $C_{123}^{\rm free}$ and also the logarithmic
terms $\l' \log|x_1|$ and $\l' \log|x_2|$ due to interactions of
the double-trace in $\bar{\cO}_{n_{1}n_{2}n_{3}}^{J}(0)$ with the
BMN operators at $x_1$ and $x_2$. But, these mixing effects cannot affect
the third logarithm, $\l' \log|x_1-x_2|$. Hence out programme is
to assume the conformal form, and by carefully evaluating the terms
proportional to $\l' \log|x_1-x_2|$, to determine $C_{123}$. In doing so we
can neglect the double-trace corrections and work with the
original single-trace expressions.

We start with
\be
 G_{3}(x_{1},x_{2})=
 \langle \bar{\cO}_{n_{1}n_{2}n_{3}}^{J}(0)
 \cO_{n_{1}' n_{2}' n_{3}'}^{J_{1}}(x_{1}) 
 \cO_{\rm vac}^{J_{2}}(x_{2})\rangle \ .
\ee
The calculation is done as in the previous subsection, except that 
now we have additional phase factors coming from non-zero
$n_{2}'$ and $n_{3}'$.
The result for phase factor $X$ coming from the free diagrams of 
Figure 5 
is obtained from \eqref{twentyone}
by substituting $n_{2}-{n_{2}'}/{y}$ for
$n_{2}$ and $n_{3}-{n_{3}'}/{y}$ for  $n_{3}$
where $y= J_1/J$.
The final result is 
\begin{eqnarray}
X=J^{3}2^{3}\, \frac{\sin(\pi n_{2}y)
\sin(\pi n_{3}y)\sin(\pi (n_{2}+n_{3})y)}{(2\pi)^{3}(
n_{2}+n_{3}-\frac{n_{3}'+n_{2}'}{y})
(n_{2}-\frac{n_{2}'}{y})(n_{3}-\frac{n_{3}'}{y})} \ .
\end{eqnarray}
To simplify notation somewhat, we will define 
\begin{eqnarray}
\Pi :=\, \frac{\sin(\pi n_{2}y)\sin(\pi n_{3}y)
\sin(\pi (n_{2}+n_{3})y)}{(n_{2}+n_{3}
-\frac{n_{3}'+n_{2}'}{y})(n_{2}
-\frac{n_{2}'}{y})(n_{3}-\frac{n_{3}'}{y})} \ .
\end{eqnarray}
Next we evaluate the interacting diagrams of Figure 6,
\begin{eqnarray}
Y_{123}=\sum_{a=1}^{J_{1}+1}\sum_{b=a+1}^{J_{1}+2}
[-\mbox{\={q}}_{2}^{J-(J_{1}+3-a-1)}
\mbox{\={q}}_{3}^{J-(J_{1}+3-b)}+\mbox{\={q}}_{2}^{J-(J_{1}+3-a-2)}
\mbox{\={q}}_{3}^{J-(J_{1}+3-b-1)}]r_{2}^{a-1}r_{3}^{b-2}
\nonumber\\
+\sum_{a=3}^{J_{1}+2}\sum_{b=a+1}^{J_{1}+3}
[\mbox{\={q}}_{2}^{a-3}\mbox{\={q}}_{3}^{b-4}
-\mbox{\={q}}_{2}^{a-2}\mbox{\={q}}_{3}^{b-3}]r_{2}^{a-3}r_{3}^{b-4}
\nonumber\\
=\sum_{a=1}^{J_{1}+1}\sum_{b=a+1}^{J_{1}+2}
\mbox{\={q}}_{2}^{-(J_{1}-a+1)}\mbox{\={q}}_{3}^{-(J_{1}-b+2)}
r_{2}^{a-1}r_{3}^{b-2}(1-\mbox{\={q}}_{2}^{-1}\mbox{\={q}}_{3}^{-1})
+\sum_{a=3}^{J_{1}+2}\sum_{b=a+1}^{J_{1}+3}
\mbox{\={q}}_{2}^{a-3}\mbox{\={q}}_{3}^{b-4}r_{2}^{a-3}
r_{3}^{b-4}(1-\mbox{\={q}}_{2}\mbox{\={q}}_{3})
\nonumber\\
\rightarrow \frac{2\pi i (n_{2}+n_{3})}{J}
\left[\sum_{a=3}^{J_{1}+2}
\sum_{b=a+1}^{J_{1}+3}(\mbox{\={q}}_{2}r_{2})^{a-3}
(\mbox{\={q}}_{3}r_{3})^{b-4}-\sum_{a=1}^{J_{1}+1}
\sum_{b=a+1}^{J_{1}+2}(\mbox{\={q}}_{2}r_{2})^{a-1}
(\mbox{\={q}}_{3}r_{3})^{b-2}\mbox{\={q}}_{2}^{-J_{1}}
\mbox{\={q}}_{3}^{-J_{1}}\right]
\end{eqnarray}
Converting the last sum into an integral we obtain,
\begin{eqnarray} 
Y_{123}={2\pi i (n_{2}+n_{3})}{J}
\int_{0}^{y}da\int_{a}^{y}db 
\, e^{-2\pi i (n_{2}-\frac{n_{2}'}{y})a} 
e^{-2\pi i (n_{3}-\frac{n_{3}'}{y})b}
[1- e^{2\pi i (n_{3}+n_{2})y}]
\nonumber\\
={2\pi i}{J}\nonumber(n_{2}+n_{3})[1- e^{2\pi i 
(n_{3}+n_{2})y}]\frac{ e^{(A_{3}+A_{2})y}A_{3}
-e^{A_{3}y}A_{2}
-e^{A_{3}y}A_{3}+A_{2}}{(A_{3}+A_{2})A_{3}A_{2}}
\end{eqnarray}
where $A_{2}=-2\pi i (n_{2}-\frac{n_{2}'}{y})$, 
$A_{3}=-2\pi i (n_{3}-\frac{n_{3}'}{y})$, 
$r_{2}=e^{2 \pi i n_{2}'/J_{1}}$ and 
$r_{3}=e^{2 \pi i n_{3}'/J_{1}} $.
Expression for $Y_{132}$ is obtained by interchanging
$A_{2}$ and $A_{3}$ in  $Y_{123}$.
 After some algebra we get
\be
Y_{123}+Y_{132}=\frac{4(n_{3}+n_{2})(n_{3}+n_{2}-
\frac{n_{2}'+n_{3}'}{y})J}{\pi}
\  \Pi
\ee

Similarly, for diagrams of Figure 7 we obtain (in the BMN limit)
\begin{eqnarray} 
Y_{231}={2\pi i n_{2}}{J}
\int_{0}^{y}da\int_{0}^{y}db 
\, e^{2\pi i (n_{2}-\frac{n_{2}'}{y})b} 
e^{2\pi i (n_{3}-\frac{n_{3}'}{y})(b-a)}
[1- e^{-2\pi i n_{2}y}]
\nonumber\\
={2\pi i n_{2}}{J}[1- e^{-2\pi i n_{2}y}]
\frac{e^{-A_{2}y}A_{2}+e^{-A_{2}y}A_{3}-
e^{-(A_{3}+A_{2})y}A_{2}-A_{3}}{(A_{3}+A_{2})A_{3}A_{2}}
\end{eqnarray}

Now consider the above diagrams with  $\phi_{1}$ and 
$\phi_{3}$ exchanged. In the BMN limit we find
\begin{eqnarray} 
Y_{213}=-{2\pi i n_{2}}{J}\int_{0}^{y}
da\int_{0}^{y}db \, e^{-2\pi i 
(n_{2}-\frac{n_{2}'}{y})(-a)} 
e^{-2\pi i (n_{3}-\frac{n_{3}'}{y})(b-a)}
[1- e^{-2\pi i n_{2}y}]
\nonumber\\
=-{2\pi i n_{2}}{J}[1- e^{-2\pi i n_{2}y}]
\frac{e^{-A_{2}y}A_{3}+e^{A_{3}y}A_{2}
-A_{2}-A_{3}}{(A_{3}+A_{2})A_{3}A_{2}}
\end{eqnarray}
Summing the two we arrive at
\be
Y_{231}+Y_{213}=\frac{4 n_{2}(n_{2}-\frac{n_{2}'}{y})J}{\pi}
 \ \Pi \ .
\ee
Similarly for the diagrams of 
Figure 8 we get:
\be
Y_{312}+Y_{321}=\frac{4 n_{3}(n_{3}-\frac{n_{3}'}{y})J}{\pi}\ \Pi \ .
\ee

Using the above phase factors and concentrating 
on the logarithmic terms of the 3-point function we have
\begin{eqnarray}
G_{3}(x_{1},x_{2})=C_{123}^{(0)} \ 
\Delta(x_{1})^{J_{1}+3}\Delta(x_{2})^{J_{2}}
[1-\frac{\lambda'}{4}(n_{2}(n_{2}-\frac{n_{2}'}{y})
+n_{3}(n_{3}-\frac{n_{3}'}{y})\nonumber\\+(n_{3}
+n_{2})(n_{3}+n_{2}-\frac{n_{2}'+n_{3}'}{y}))
\log \frac{x_{1}^{2}x_{2}^{2}}{x_{12}^{2}}
+C'\log x_{1}^{2}+...]
\end{eqnarray}
The last term in the equation above comes from the diagrams of  
Figure 9 which no longer sum to zero as 
in the case with two supergravity states. 
However we do not need to know the coefficient of this term since the 
$\log x_{1}^{2}$ receives corrections from the the double-trace operators.
From the equation above one can easily read 
the coefficient $C_{123}$ at order $g_{2}$
\be \label{cst123}
 C_{123}=\, C_{123}^{(0)}\, \frac{n_{2}(n_{2}
 -\frac{n_{2}'}{y})+n_{3}(n_{3}
-\frac{n_{3}'}{y})+(n_{3}+n_{2})(n_{3}+n_{2}
-\frac{n_{2}'+n_{3}'}{y})}{
n_{2}^{2}+n_{3}^{2}+(n_{2}+n_{3})^{2}
-\frac{n_{2}'^{2}+n_{3}'^{2}+(n_{2}'+n_{3}')^{2}}
{y^{2}}
}
\ee
where 
\be \label{c0123}
C_{123}^{(0)}=\frac{J\sqrt{J}\sqrt{1-y}}{N y \pi^{3}}
\, \frac{\sin(\pi n_{2}{y})\sin(\pi n_{3}{y})
\sin(\pi (n_{2}+n_{3}){y})}{(n_{2}+n_{3}
-\frac{n_{3}'+n_{2}'}{y})(n_{2}
-\frac{n_{2}'}{y})(n_{3}-\frac{n_{3}'}{y})}
\ee

\medskip
 
 {\it Three-point correlator $ G'_{3}(x_{1},x_{2})$}

Finally we consider the $G'_{3}$ function with two string states   
\be
 G'_{3}(x_{1},x_{2})=
 \langle\bar{\cO}_{n_{1}n_{2}n_{3}}^{J}(0)
 \cO_{n_{2}' -n_{2}'}^{J_{1}}(x_{1}) \cO_{0}^{J_{2}}(x_{2})> 
\ee
Similarly to the earlier analysis, we determine $P$ from 
free diagrams in Figure 10,
\be
P=- \frac{J^{3}\sin(\pi n_{2}y)
\sin(\pi n_{3}y)\sin(\pi(n_{3}
+n_{2}) y)}{\pi^{3}(n_{3}+n_{2}
-\frac{n_{2}'}{y})(n_{2}-\frac{n_{2}'}{y})n_{3}}
\ee
The diagrams of Figure 11, Figure 12 and Figure 12 
lead to the expressions for $Q_{1}$, $Q_{2}$ and $Q_{3}$ 
respectively
\begin{eqnarray}
Q_{1}=-\frac{4}{\pi}J(n_{3}+n_{2})
(n_{3}+n_{2}-\frac{n_{2}'}{y})\,
\frac{\sin(\pi n_{2}y)
\sin(\pi n_{3}y)\sin(\pi(n_{3}+n_{2}) y)}{(n_{3}+n_{2}
-\frac{n_{2}'}{y})(n_{2}-\frac{n_{2}'}{y})n_{3}}
\end{eqnarray}
\begin{eqnarray}
Q_{2}=-\frac{4}{\pi}Jn_{2}\left(n_{2}-\frac{n_{2}'}{y}\right)\,
\frac{\sin(\pi n_{2}y)
\sin(\pi n_{3}y)\sin(\pi(n_{3}+n_{2}) y)}{(n_{3}+n_{2}
-\frac{n_{2}'}{y})(n_{2}-\frac{n_{2}'}{y})n_{3}}
\end{eqnarray}
\begin{eqnarray}
Q_{3}=-\frac{4}{\pi}Jn_{3}^{2}\,
\frac{\sin(\pi n_{2}y)
\sin(\pi n_{3}y)\sin(\pi(n_{3}+n_{2}) 
y)}{(n_{3}+n_{2}
-\frac{n_{2}'}{y})(n_{2}-\frac{n_{2}'}{y})n_{3}}
\end{eqnarray} 
Using the above results the three-point function reads
\begin{eqnarray}
G_{3}'(x_{1},x_{2})=\tilde{C}_{123}^{(0)}\ 
\Delta(x_{1})^{J_{1}+2}\Delta(x_{2})^{J_{2}+1}
[1-\frac{\lambda'}{4}(n_{3}^{2}+n_{2}(n_{2}
-\frac{n_{2}'}{y})
\\
+(n_{3}
+n_{2})(n_{3}+n_{2}-\frac{n_{2}'}{y}))
\log \frac{x_{1}^{2}x_{2}^{2}}{x_{12}^{2}}
+C'\log x_{1}^{2}+C''\log x_{2}^{2}+...]
\nonumber
\end{eqnarray}
where 
\be
\label{ctil0}
\tilde{C}_{123}^{(0)}=-\frac{J\sqrt{J}}{\sqrt{y} 
N \pi^{3}}\frac{\sin(\pi n_{2}y)
\sin(\pi n_{3}y)\sin(\pi(n_{3}+n_{2}) 
y)}{(n_{3}+n_{2}-\frac{n_{2}'}{y})
(n_{2}-\frac{n_{2}'}{y})n_{3}}
\ee
And our final expression for $\tilde{C}_{123}$ is
\be
\label{ctil123}
\tilde{C}_{123}=\tilde{C}_{123}^{(0)}
\frac{n_{3}^{2}+n_{2}(n_{2}-\frac{n_{2}'}{y})
+(n_{3}+n_{2})(n_{3}+n_{2}-\frac{n_{2}'}{y})}
{n_{2}^{2}+ n_{3}^{2}+(n_{2}+n_{3})^{2}
-{\frac{2n_{2}'^{2}}{x^{2}}}}
\ee

\section{Tests of the correspondence}

In this section we test the correspondence proposed in 
\cite{CK} and outlined in Section 2 against the gauge theory
results of Section 4.

\subsection{ $\bf 1_{-(n_{2}'+n_{3}')} 2_{n_{2}'} 3_{n_{3}'} +   
vac \to 1_{-(n_{2}+n_{3})} 2_{n_{2}} 3_{n_{3}}$}

In this case the external string state is
\begin{eqnarray}
\langle\Phi|=\langle0|\alpha^{3 i_{1}}_{-(n_{2}+n_{3})}
\alpha^{3 i_{2}}_{n_{2}}\alpha^{3 i_{3}}_{n_{3}}
\alpha^{1 i_{1}}_{-(n_{2}'+n_{3}')}
\alpha^{1 i_{2}}_{n_{2}'}\alpha^{1 i_{3}}_{n_{3}'}
\end{eqnarray}
In the expression above, 1 and 3 denote the first and the third
string in the vertex. In order to avoid confusion, and distinguish
from the indices  corresponding to scalar impurites, 
$\phi_1$, $\phi_2$, $\phi_3$,
we label the latter with $i_1$, $i_2$, $i_3$.
Since all three SYM impurities are different, we note
that $i_1 \neq i_2 \neq i_3$ and the repeated indices
in the external states are not summed over.

The contribution of the first part of the prefactor ${\sf P}_{I}$ is 
\begin{eqnarray}
\langle\Phi| {\sf P}_{I} \ket{V_B}
=&&\frac{1}{2\mu}
\left[\frac{n_{2}'^{2}+n_{3}'^{2}+
(n_{2}'+n_{3}')^{2}}{y^{2}}-n_{2}^{2}-n_{3}^{2}
-(n_{2}+n_{3})^{2}\right] \nonumber\\
&&\times \hat{N}^{31}_{n_{2},n_{2}'}
\hat{N}^{31}_{n_{3},n_{3}'}\hat{N}^{31}_{n_{2}+n_{3},n_{2}'+n_{3}'}
\end{eqnarray}
which, using the expression for the Neumann matrices 
(see Appendix),
\be \label{Nppex}
\Nh^{31}_{mn} =\Nh^{31}_{-m-n}  = \frac{ (-1)^{m+n+1}}{\pi}
\frac{\sin(\pi m y)}{\sqrt{y}(m-n/y)}+\cO\left(
\frac{1}{\mu^2}\right),
\ee
becomes
\be
\langle\Phi| {\sf P}_{I} \ket{V_B}
=-\frac{1}{2\mu y^{3/2} \pi^{3}}
\left[\frac{n_{2}'^{2}+n_{3}'^{2}+(n_{2}'
+n_{3}')^{2}}{y^{2}}-n_{2}^{2}-n_{3}^{2}-(n_{2}+n_{3})^{2}\right]
\ {\Pi} 
\ee
Now consider  ${\sf P}_{II}$. The contributing terms in 
${\sf P}_{II}$ for the state under consideration are 
\begin{eqnarray}
{\sf P}_{II}=\frac{1}{2}[(\frac{\omega_{3n_{2}}}{\alpha_{3}}
+\frac{\omega_{1n_{2}'}}{\alpha_{1}})(\hat{N}^{31}_{n_{2},-n_{2}'}
-\hat{N}^{31}_{n_{2},n_{2}'})
\alpha^{3 i_{2}\dagger}_{n_{2}}\alpha^{1 i_{2}\dagger}_{n_{2}'}
\nonumber\\
+(\frac{\omega_{3n_{3}}}{\alpha_{3}}
+\frac{\omega_{1n_{3}'}}{\alpha_{1}})(\hat{N}^{31}_{n_{3},-n_{3}'}
-\hat{N}^{31}_{n_{3},n_{3}'})\alpha^{3 i_{3}\dagger}_{n_{3}}
\alpha^{1 i_{3}\dagger}_{n_{3}'}
\nonumber\\
+(\frac{\omega_{3n_{1}}}{\alpha_{3}}
+\frac{\omega_{1n_{1}'}}{\alpha_{1}})
(\hat{N}^{31}_{-n_{1},n_{1}'}-
\hat{N}^{31}_{-n_{1},-n_{1}'})\alpha^{3 i_{1}\dagger}_{n_{1}}
\alpha^{1 i_{1}\dagger}_{n_{1}'}]
\end{eqnarray}
Using this 
 we find
\begin{eqnarray}
\langle\Phi| {\sf P}_{II}\ket{V_B}=
-\frac{1}{4\mu}[(n_{2}^{2}
-\frac{n_{2}'^{2}}{y^{2}})N^{31}_{-n_{2},-n_{2}'}
\hat{N}^{31}_{n_{3},n_{3}'}\hat{N}^{31}_{-n_{1},-n_{1}'}
\nonumber\\
+(n_{3}^{2}-\frac{n_{3}'^{2}}{y^{2}})N^{31}_{-n_{3},-n_{3}'}
\hat{N}^{31}_{n_{2},n_{2}'}\hat{N}^{31}_{-n_{1},-n_{1}'}
\nonumber\\
+(n_{1}^{2}-\frac{n_{1}'^{2}}{y^{2}})N^{31}_{n_{1},n_{1}'}
\hat{N}^{31}_{n_{3},n_{3}'}\hat{N}^{31}_{n_{2},n_{2}'}]
\end{eqnarray}
Recalling that $n_1=-(n_2+n_3)$ and 
$n_{1}'=-(n_{2}'+n_{3}')$ and substituting into
the above result the expressions \eqref{Nppex} for the
Neumann matrices $\Nh$ and expressions 
for the Neumann matrices $N_{-,-}$ (see Appendix)
\be
N^{31}_{-m-n} = \frac{2 (-1)^{m+n}}{\pi}
\frac{n \sin(\pi m y)}{y^{3/2}(m^2-n^2/y^2)}+\cO\left(\frac{1}{\mu^2}\right), 
\ee
we obtain
\be
\langle\Phi| P_{II}\hat{V}|0\rangle
=-\frac{1}{2\mu \pi^{3}y^{5/2}}\left[n_{2}'(n_{2}
-\frac{n_{2}'}{y})+n_{3}'(n_{3}
-\frac{n_{3}'}{y})+n_{1}'(n_{1}-\frac{n_{1}'}{y})\right]\ \Pi
\ee
Adding these results for  ${\sf P}_{I}$ and  
${\sf P}_{II}$ 
and multiplying by 
$(-1)^p C_{norm}=-g_{2}\frac{\sqrt{y(1-y)}}{\sqrt{J}}$ 
we get the string theory answer
\begin{eqnarray}
-\, \frac{g_2}{2\mu}\frac{\sqrt{1-y}}{\pi^{3}\sqrt{J}\,y}
\left[n_{2}(n_{2}-\frac{n_{2}'}{y})+n_{3}(n_{3}
-\frac{n_{3}'}{y})+(n_{3}+n_{2})(n_{3}+n_{2}
-\frac{n_{2}'+n_{3}'}{y})\right] \ \Pi 
\end{eqnarray}
which is {\it exactly the SYM result}, 
$\mu(\Delta_{1}+\Delta_{2}-\Delta_{3})C_{123}$,
since
\be \label{dimsdef}
\mu(\Delta_{1}+\Delta_{2}-\Delta_{3})=
\frac{1}{2\mu}
\left[\frac{n_{2}'^{2}+n_{3}'^{2}+
(n_{2}'+n_{3}')^{2}}{y^{2}}-n_{2}^{2}-n_{3}^{2}
-(n_{2}+n_{3})^{2}\right]
\ee
and $C_{123}$, given by \eqref{cst123}, \eqref{c0123},
reads
 \be \label{cstagain}
 C_{123}=\, 
 \frac{J\sqrt{J}\sqrt{1-y}}{N y \pi^{3}}
 \, \frac{n_{2}(n_{2}
 -\frac{n_{2}'}{y})+n_{3}(n_{3}
-\frac{n_{3}'}{y})+(n_{3}+n_{2})(n_{3}+n_{2}
-\frac{n_{2}'+n_{3}'}{y})}{
n_{2}^{2}+n_{3}^{2}+(n_{2}+n_{3})^{2}
-\frac{n_{2}'^{2}+n_{3}'^{2}+(n_{2}'+n_{3}')^{2}}
{y^{2}}
} \ \Pi.
\ee

\bigskip

\subsection{$\bf 1_{n_{2}'} 
2_{-(n_{2}'+n_{3}')} 3_{n_{3}'} + vac
\to   1_{-(n_{2}+n_{3})}   2_{n_{2}} 3_{n_{3}}$}

In this case the external string state is
\begin{eqnarray}
\langle\Phi|=\langle0|\alpha^{3 i_{1}}_{-(n_{2}
+n_{3})}\alpha^{3 i_{2}}_{n_{2}}\alpha^{3 i_{3}
}_{n_{3}}\alpha^{1 i_{1}}_{n_{2}'}\alpha^{1 i_{2}
}_{-(n_{2}'+n_{3}')}\alpha^{1 i_{3}}_{n_{3}'}
\end{eqnarray}
and the operator $\cO_{1}(x_{1})$ is
\begin{eqnarray}
\cO_{1}(x_{1})=\frac{1}{J\sqrt{N^{J+3}}}\sum_{0\leq a,b
}^{a+b\leq {J}}[ r_{2}^{a}r_{3}^{a+b} 
\tr(\phi_{2}Z^{a}\phi_{1}Z^{b}\phi_{3}Z^{J-a-b})\nonumber\\
+ r_{3}^{a}r_{2}^{a+b} 
\tr(\phi_{2}Z^{a}\phi_{3}Z^{b}\phi_{1}Z^{J-a-b})]
\end{eqnarray}
Using the cyclic property of the trace,  
$\cO_{1}(x_{1})$ can also be written as
\begin{eqnarray}
\cO_{1}(x_{1})=\frac{1}{J\sqrt{N^{J+3}}}
\sum_{0\leq a,b}^{a+b\leq {J}}
[\tilde{ r_{2}}^{a}\tilde{ r_{3}}^{a+b}
\tr(\phi_{1}Z^{a}\phi_{2}Z^{b}\phi_{3}Z^{J-a-b})
\nonumber\\+\tilde{ r_{3}}^{a}\tilde{ r_{2}}^{a+b} 
\tr(\phi_{1}Z^{a}\phi_{3}Z^{b}\phi_{2}Z^{J-a-b})]
\end{eqnarray}
where $\tilde{ r_{2}}= (r_{2} r_{3})^{-1}$ 
and $\tilde{ r_{3}}=r_{3}$. 
Hence, we find the 3-point function 
for this process 
by substituting 
$n_{2}'\to -(n_{2}'+n_{3}') $, $n_{3}'\to n_{3}' $ 
and $-(n_{2}'+n_{3}')=n_{1} \to n_{2}'$ into the SYM 
expression \eqref{cstagain} and also to \eqref{dimsdef}.

Thus, the result on the gauge theory 
side for this process is
\begin{eqnarray}
\mu(\Delta_{1}+\Delta_{2}-\Delta_{3})C_{123}=
-\, \frac{g_2}{2\mu}\frac{\sqrt{1-y}}{\pi^{3}\sqrt{J}\,y}
\frac{\sin(\pi n_{2} y)\sin(\pi n_{3} y)
\sin(\pi (n_{2}+ n_{3}) y)}{(n_{2}+
\frac{n_{2}'+n_{3}'}{y})(n_{3}
-\frac{n_{3}'}{y})(n_{2}+n_{3}+\frac{n_{2}'}{y})}
\nonumber\\
\times \left[n_{2}(n_{2}+\frac{n_{2}'+n_{3}'}{y})
+n_{3}(n_{3}-\frac{n_{3}'}{y})
+(n_{2}+n_{3})(n_{2}+n_{3}+\frac{n_{2}'}{y})\right]
\label{sevf}
\end{eqnarray}

On the string theory side of the correspondence
the calculation follows the same lines as in {\bf 5.1},
and the result
is in precise agreement with the SYM formula \eqref{sevf}.

This is also true for the remaining four cases 
involving different permutations of the three 
$\phi$'s in the trace of the shortest BMN operator.

\subsection{$\bf 1_{-n_{2}'} 2_{n_{2}'} + 3_{0} \to 
1_{-(n_{2}+n_{3})}   2_{n_{2}}   3_{n_{3}}$}

Here we consider the case which corresponds
to $G_3'$, i.e. where instead of the vacuum 
state for the string 2, we have a supergravity state. 
The external string state now is
\begin{eqnarray}
\langle\Phi|=\langle0|\alpha^{3 i_{1}}_{-(n_{2}+n_{3})}
\alpha^{3 i_{2}}_{n_{2}}\alpha^{3 i_{3}}_{n_{3}}
\alpha^{1 i_{1}}_{-n_{2}'}\alpha^{1 i_{2}}_{+n_{2}'}
\alpha^{2 i_{3}}_{0}
\end{eqnarray}
The result in the SYM side can be obtained from 
\eqref{ctil123} and \eqref{ctil0} and is
\begin{eqnarray}
\mu(\Delta_{1}+\Delta_{2}-\Delta_{3})\tilde{C}_{123}
=\frac{1}{2\mu}\frac{J\sqrt{J}}{\sqrt{y} 
N \pi^{3}}\frac{\sin(\pi n_{2}{y})
\sin(\pi n_{3}{y})\sin(\pi(n_{3}
+n_{2}) {y})}{(n_{3}+n_{2}-\frac{n_{2}'}{y})
(n_{2}-\frac{n_{2}'}{y})n_{3}}
\nonumber\\
(n_{3}^{2}+n_{2}(n_{2}-\frac{n_{2}'}{y})
+(n_{3}+n_{2})(n_{3}+n_{2}-\frac{n_{2}'}{y})).
\end{eqnarray}
This expression is again, as it is easy to check
using the Neumann matrices from the Appendix,
in precise agreement with the expression on
the string theory side  of the duality relation
\eqref{hhhh}.

\subsection{$\bf 1_{n_{2}'} 2_{-n_{2}'}  + 3_{0}   
\to 1_{-(n_{2}+n_{3})}   2_{n_{2}}   3_{n_{3}}$}

As a final example we consider the case where  
the external string state  is
\begin{eqnarray}
\langle\Phi|=\langle0|\alpha^{3 i_{1}}_{-(n_{2}+n_{3})}
\alpha^{3 i_{2}}_{n_{2}}\alpha^{3 i_{3}}_{n_{3}}
\alpha^{1 i_{1}}_{n_{2}'}\alpha^{1 i_{2}}_{-n_{2}'}
\alpha^{2 i_{3}}_{0}
\end{eqnarray}
The result on the SYM side 
is given by 
\begin{eqnarray}
\mu(\Delta_{1}+\Delta_{2}-\Delta_{3})
\tilde{C}_{123}=\frac{1}{2\mu}\frac{J\sqrt{J}}
{\sqrt{y} N \pi^{3}}\frac{\sin(\pi n_{2}{y})
\sin(\pi n_{3}{y})
\sin(\pi(n_{3}+n_{2}) {y})}{(n_{3}+n_{2}
+\frac{n_{2}'}{y})(n_{2}
+\frac{n_{2}'}{y})n_{3}}\nonumber\\
(n_{3}^{2}+n_{2}(n_{2}+\frac{n_{2}'}{y})
+(n_{3}+n_{2})(n_{3}+n_{2}+\frac{n_{2}'}{y}))
\end{eqnarray}
which is in precise agreement with the string vertex--correlator 
duality prediction \eqref{hhhh}.

\bigskip
\bigskip

\section*{Acknowledgements} 

We thank Chong-Sun Chu and Gabriele Travaglini for 
many useful discussions. GG would like to thank Konstantina Mara
for her support. He also acknowledges grant from the State Scholarship
Foundation of Greece (I.K.Y.).

\newpage

\section*{Appendix}

Here we outline the pp-wave string field theory conventions used in the text.
The combination $\a'p^+$ for the r-th string is denoted $\a_\rr$
and $\sum_{\rr=1}^3 \a_\rr =0$. As is standard in
the literature, we will choose a frame in which $\a_3=-1$
\be \label{frame}
\a_\rr = \a'p^+_{(\rr)} \, : \qquad \a_3=-1, \qquad \a_1=y, \qquad \a_2=1-y.
\ee
In terms of the $U(1)$ R-charges 
of the BMN operators in the 
three-point function, $\langle \cO_1^{J_1} \cO_2^{J_2} \bar\cO_3^{J}
\rangle$, 
where $J=J_1+J_2$, we have
\be
y=\frac{J_1}{J}, \qquad 1-y=\frac{J_2}{J}, \qquad 0<y<1.
\ee
The effective SYM coupling constant \eqref{lampr} in the frame \eqref{frame}
takes a simple form
\be \label{lamprsim}
 \l' =  \frac{1}{(\mu p^+ \a')^2}\ \equiv \frac{1}{(\mu \a_3)^2}\ 
= \frac{1}{\mu^2}.
\ee
Here $\mu$ is the mass parameter which appears in the pp-wave metric, in
the chosen frame it is dimensionless\footnote{It is $p^{+}\mu$ which is
invariant under longitudinal boosts and is frame-independent.} and the
expansion in powers of $1/\mu^2$ is equivalent to the perturbative
expansion in $\l'$. Finally the frequencies are defined via,
\be
\omega_{\rr m}= \sqrt{m^2+(\mu\a_\rr)^2}.
\ee
 
The infinite-dimensional Neumann matrices, $N_{mn}^{\rr\rs}$ are usually specified
in the original $a$-oscillator basis of the string field theory. 
In this basis
the bosonic overlap factor $\ket{V_B}$ of the 3-strings vertex is given by
\be \label{Vbab}
\ket{V_B} = \exp( \frac{1}{2} \sum_{\rr,\rs=1}^3 \sum_{m,n=-\infty}^\infty
 a^{\rr\, I\dag}_{m} N_{mn}^{\rr\rs}  a^{\rs\, I\dag}_{n}) \ket{0}.
\ee
However, for the purposes of the pp-wave/SYM correspondence it is more convenient
to use another, the so-called BMN or $\a$-basis of string oscillators,
which is in direct correspondence with the BMN operators in gauge theory. The two
bases are related as follows:
\be \label{bmn-a}
\a_n = \frac{1}{\sqrt{2}}(a_{|n|} - i \, {\rm sign}(n) a_{-|n|}),
\quad \a_0 = a_0,
\ee
and satisfy the same oscillator algebra
\be
[\a_m, \a_n^\dag] = \d_{mn}.
\ee
In this basis, the bosonic overlap factor \eqref{Vbab} in the vertex reads
\be \label{Vbal}
\ket{V_B} = \exp( \frac{1}{2} \sum_{\rr,\rs=1}^3 \sum_{m,n=-\infty}^\infty
\a^{\rr\, I\dag}_{m} \hat{N}_{mn}^{\rr\rs}  \a^{\rs\, I\dag}_{n}),
\ee
where $\Nh$ are the Neumann matrices in the $\a$-basis and
are related to the $N$'s via (here $m,n>0$):
\bea 
&& \Nh^{\rr\rs}_{mn} = \Nh^{\rr\rs}_{-m-n} := 
\frac{1}{2}(N^{\rr\rs}_{mn} -N^{\rr\rs}_{-m-n}), \\
&& \Nh^{\rr\rs}_{m-n} = \Nh^{\rr\rs}_{-m n} := 
\frac{1}{2}(N^{\rr\rs}_{mn} + N^{\rr\rs}_{-m-n}), \\
&& \Nh^{\rr\rs}_{m0} =\Nh^{\rr\rs}_{-m0} := \frac{1}{\sqrt{2}} 
N^{\rr\rs}_{m0}=
\Nh^{\rr\rs}_{0m} =\Nh^{\rr\rs}_{0-m}, \label{poszer}\\
&&  \Nh^{\rr\rs}_{00}:=N^{\rr\rs}_{00}.
\eea

We now copy the explicit expressions for the Neumann matrices 
\cite{HSSV}
in the $a$-basis from the Appendix of \cite{CK}.
These expressions are needed
for our calculations in Section 5.
\bea
&&N^{31}_{mn} = \frac{2 (-1)^{m+n+1}}{\pi}
\frac{m \sin(\pi m y)}{\sqrt{y}(m^2-n^2/y^2)}+\cO\left(
\frac{1}{\mu^2}\right), \\
&&N^{32}_{mn} = \frac{2 (-1)^{m}}{\pi}
\frac{m \sin(\pi m y)}{\sqrt{1-y}(m^2-n^2/(1-y)^2)}+\cO\left(
\frac{1}{\mu^2}\right), \\
&&N^{21}_{mn} = \frac{1}{\mu}\,\frac{ (-1)^{n+1}}{2\pi}
\frac{1}{\sqrt{y(1-y)}}+\cO\left(\frac{1}{\mu^3}\right), \\
&&N^{33}_{mn} = \cO\left(\frac{1}{\mu^3}\right), \\
&&N^{11}_{mn} = \frac{1}{\mu}\,\frac{ (-1)^{m+n}}{2\pi}
\frac{1}{y}+\cO\left(\frac{1}{\mu^3}\right), \\
&&N^{22}_{mn} = \frac{1}{\mu}\,\frac{ 1}{2\pi}
\frac{1}{1-y}+\cO\left(\frac{1}{\mu^3}\right).
\eea

\bea
&&N^{31}_{-m-n} = \frac{2 (-1)^{m+n}}{\pi}
\frac{n \sin(\pi m y)}{y^{3/2}(m^2-n^2/y^2)}+\cO\left(\frac{1}{\mu^2}\right), \\
&&N^{32}_{-m-n} = \frac{2 (-1)^{m+1}}{\pi}
\frac{n \sin(\pi m y)}{(1-y)^{3/2}(m^2-n^2/(1-y)^2)}+\cO\left(
\frac{1}{\mu^2}\right), \\
&&N^{21}_{-m-n} = \cO\left(\frac{1}{\mu^3}\right), \\
&&N^{33}_{-m-n} = \frac{1}{\mu}\,
\frac{2(-1)^{m+n}}{\pi} \sin(\pi m y)\sin(\pi n y)
+ \cO\left(\frac{1}{\mu^3}\right), \\
&&N^{11}_{-m-n} = \cO\left(\frac{1}{\mu^3}\right), \\
&&N^{22}_{-m-n} = \cO\left(\frac{1}{\mu^3}\right).
\eea

\bea
&&N^{33}_{00} = 0, \quad N^{31}_{00} =-\sqrt{y}, \quad 
N^{32}_{00} =-\sqrt{1-y}, \\
&&N^{12}_{00} = \frac{1}{\mu}\,\frac{(-1)}{4\pi}\frac{1}{\sqrt{y(1-y)}}  
=\, N^{21}_{00}, \\
&&N^{11}_{00} = \frac{1}{\mu}\,\frac{1}{4\pi}
\frac{1}{y}, \\
&&N^{22}_{00} = \frac{1}{\mu}\,\frac{1}{4\pi}
\frac{1}{1-y}.
\eea
For the zero-positive Neumann matrices below we have 
\be
N^{31}_{0n}= 0, \quad N^{32 }_{0n} =0, \quad N^{33 }_{0n} =0.
\ee
\bea
&&N^{13}_{0n} = \frac{\sqrt{2} (-1)^{n+1}}{\pi}
\frac{\sin(\pi n y)}{n\sqrt{y}}+\cO\left(\frac{1}{\mu^2}\right), \\
&&N^{23}_{0n} = \frac{\sqrt{2} (-1)^{n}}{\pi}
\frac{\sin(\pi n y)}{n\sqrt{1-y}}+\cO\left(\frac{1}{\mu^2}\right), \\
&&N^{21}_{0n} = \frac{1}{\mu}\,\frac{\sqrt{2}(-1)^{n+1} }{4\pi}
\frac{1}{ \sqrt{y(1-y)}}+\cO\left(\frac{1}{\mu^3}\right), \\
&&N^{12}_{0n} = -\frac{1}{\mu}\,\frac{\sqrt{2}}{4\pi}
\frac{1}{\sqrt{y(1-y)}}+\cO\left(\frac{1}{\mu^3}\right), \\
&&N^{11}_{0n} = \frac{1}{\mu}\,\frac{\sqrt{2} (-1)^{n}}{4\pi}
\frac{1}{y}+\cO\left(\frac{1}{\mu^3}\right), \\
&&N^{22}_{0n} = \frac{1}{\mu}\,\frac{\sqrt{2}}{4\pi}
\frac{1}{1-y}+\cO\left(\frac{1}{\mu^3}\right).
\eea

\end{document}